\documentclass[12pt]{article}

\usepackage{amsfonts,amsmath,amssymb}
\usepackage{enumerate}
\usepackage{hyperref}
\newcommand{\be}{\begin{equation}}
\newcommand{\ee}{\end{equation}}

\newcommand{\bea}{\begin{eqnarray}}
\newcommand{\eea}{\end{eqnarray}}

\newcommand{\bes}{\begin{subequations}}
\newcommand{\ees}{\end{subequations}}

\newcommand{\nn}{\nonumber\\}

\newcommand{\cA}{{\cal A}}
\newcommand{\Tr}{\mbox{Tr}}

\newcommand{\ep}{\epsilon}
\newcommand{\w}{\wedge}

\let\l=\lambda  \let\n=\nu

\def\nn{\nonumber\\} \def\bd{\begin{document}} \def\ed{\end{document}}
\def\ds{\documentstyle} \let\fr=\frac \let\bl=\bigl \let\br=\bigr
\let\Br=\Bigr \let\Bl=\Bigl
\let\bm=\bibitem
\let\na=\nabla
\def\ba{\begin{array}}
\def\ea{\end{array}}
\def\ft#1#2{{\textstyle{{\scriptstyle #1}\over {\scriptstyle #2}}}}
\def\fft#1#2{{#1 \over #2}}
\def\del{\partial}
\def\vp{\varphi}
\def\sst#1{{\scriptscriptstyle #1}}
\def\oneone{\rlap 1\mkern4mu{\rm l}}
\def\td{\tilde}
\def\wtd{\widetilde}
\def\ie{\rm i.e.\ }
\def\dalemb#1#2{{\vbox{\hrule height .#2pt
        \hbox{\vrule width.#2pt height#1pt \kern#1pt
                \vrule width.#2pt}
        \hrule height.#2pt}}}
\def\square{\mathord{\dalemb{6.8}{7}\hbox{\hskip1pt}}}
\newcommand{\ho}[1]{$\, ^{#1}$}
\newcommand{\hoch}[1]{$\, ^{#1}$}
\newcommand{\lra}{\longrightarrow}
\newcommand{\Lra}{\Leftrightarrow}
\newcommand{\ap}{\alpha^\prime}
\newcommand{\bp}{\tilde \beta^\prime}
\def\0{{\sst{(0)}}}
\def\1{{\sst{(1)}}}
\def\2{{\sst{(2)}}}
\def\3{{\sst{(3)}}}
\def\4{{\sst{(4)}}}
\def\5{{\sst{(5)}}}
\def\6{{\sst{(6)}}}
\def\7{{\sst{(7)}}}
\def\8{{\sst{(8)}}}
\def\n{{\sst{(n)}}}
\def\cA{{{\cal A}}}
\def\cF{{{\cal F}}}
\def\tV{\widetilde V}
\def\tW{\widetilde W}
\def\tH{\widetilde H}
\def\tE{\widetilde E}
\def\tF{\widetilde F}
\def\tA{\widetilde A}
\def\im{{{\rm i}}}
\def\tY{{{\wtd Y}}}
\def\ep{{\epsilon}}
\def\vep{{\varepsilon}}
\def\R{\rlap{\rm I}\mkern3mu{\rm R}}
\def\bD{{{\bar D}}}

\newcommand{\ab}{a}
\newcommand{\cT}{{\cal T}}
\newcommand{\zz}{\zeta}

\begin{document}

\makeatletter
\renewcommand{\theequation}{\thesection.\arabic{equation}}
\@addtoreset{equation}{section}
\makeatother

\baselineskip 18pt

\begin{titlepage}

\vfill

\begin{flushright}
Imperial/TP/2010/JG/03\\
AEI-2010-147 \\
\end{flushright}

\vfill

\begin{center}
   \baselineskip=16pt
   \begin{Large}\textbf{
        Wrapped M5-branes, consistent truncations\\*[5pt] and AdS/CMT}
   \end{Large}
   \vskip 1.5cm
    Aristomenis Donos$^1$, Jerome P. Gauntlett$^1$,  Nakwoo Kim$^2$ and Oscar Varela$^3$ \\
   \vskip .6cm
     \begin{small}
  \textit{$^1$Blackett Laboratory,
        Imperial College\\ London, SW7 2AZ, U.K.}
        \end{small}\\*[.4cm]
      \begin{small}
      \textit{$^2$Department of Physics and Research Institute of Basic Science \\
        Kyung Hee University, Seoul 130-701, Korea}
        \end{small}\\*[.4cm]
        \begin{small}
      \textit{$^3$ AEI, Max-Planck-Institut f\"ur Gravitationsphysik, \\
        Am M\"uhlenberg 1, D-14476 Potsdam, Germany }
        \end{small}

   \end{center}

\vfill

\begin{center}
\textbf{Abstract}
\end{center}

\begin{quote}
At the level of the bosonic fields,
we construct consistent Kaluza--Klein reductions of $D=11$ supergravity
on $\Sigma_3\times S^4$, where $\Sigma_3=H^3/\Gamma,S^3/\Gamma$ or $R^3/\Gamma$ where
$\Gamma$ is a discrete group of isometries.
 The result is the bosonic content of
an $N=2$ $D=4$ gauged supergravity with a single vector multiplet and two hypermultiplets,
whose scalar fields parametrise $SU(1,1)/U(1)\times G_{2(2)}/SO(4)$.
When $\Sigma_3=H^3/\Gamma$ the $D=4$ theory has an $AdS_4$ vacuum which uplifts to
the known supersymmetric $AdS_4\times H^3/\Gamma\times S^4$ solution of $D=11$
supergravity that describes the $N=2$ $d=3$ SCFT arising when
M5-branes wrap SLag 3-cycles $H^3/\Gamma$ in Calabi-Yau three-folds.
We use the KK reduction for $\Sigma_3=H^3/\Gamma$ to construct $D=11$
black hole solutions that describe these $d=3$ SCFTs at finite temperature and charge density and show
that there is a superconducting instability involving a charged scalar field,
and another instability involving
involving neutral fields including both scalar and vector fields.
We also use this KK reduction to construct a $D=11$ Lifshitz solution that is dual to
a $d=3$ field theory with dynamical exponent $z\sim 39$.
\end{quote}

\vfill

\end{titlepage}

\section{Introduction}

The quantum field theories arising on M5-branes are an interesting prediction
of string/M-theory. While there are many aspect of these field theories that are still poorly understood,
in the limit of a large number of M5-branes they have a good description in terms of $D=11$ supergravity
via the AdS/CFT correspondence.
In the simplest setting of coincident planar M5-branes, there is a $d=6$
maximally supersymmetric CFT which is holographically dual to $AdS_7\times S^4$.

The AdS/CFT correspondence can also be used to study the supersymmetric field theories
arising when M5-branes wrap supersymmetric cycles.
Recall that a probe M5-brane can wrap a calibrated-cycle $\Sigma_p$ in special holonomy manifolds and preserve supersymmetry.
On length scales much larger than that of the characteristic size of $\Sigma_p$ we expect a decoupled
lower-dimensional supersymmetric  field theory on the unwrapped part of the M5-brane.
In special circumstances the AdS/CFT correspondence can again be used to analyse these field theories as first
discussed by Maldacena and Nunez \cite{Maldacena:2000mw}. In particular, for certain $\Sigma_p$,
solutions of $D=11$ supergravity can be constructed that describe a holographic flow ``across-dimensions" from the M5-brane field theory
on $\mathbb{R}^{1,5-p}\times \Sigma_p$ (with suitable $R$-symmetry currents switched on)
down to a $d=6-p$ SCFT field theory on $\mathbb{R}^{1,5-p}$ which is dual to
an $AdS_{7-p}\times\Sigma_p\times S^4$ solution of $D=11$ supergravity (suitably warped and twisted)\footnote{Note that for some $\Sigma_p$ there are also solutions
which do not flow to AdS solutions in the IR as we will shortly recall.}.
Such solutions describing M5-branes wrapping holomorphic 2-cycles were considered in \cite{Maldacena:2000mw} and generalised
to M5-branes wrapping other calibrated cycles
in \cite{Acharya:2000mu}-\cite{Gauntlett:2001jj} (see \cite{Gauntlett:2003di} for a review).
In all cases the solutions were first constructed in $D=7$ gauged supergravity and then uplifted to
$D=11$ on an $S^4$.

The example that is of most interest to this paper is when M5-branes wrap special Lagrangian (SLag) 3-cycles
in Calabi-Yau three-folds. In this case, at large distances,
one expects a $d=3$ quantum field theory with $N=2$ supersymmetry.
When $\Sigma=H^3/\Gamma$, where $H^3$ is hyperbolic three-space and
$\Gamma$ is a freely acting discrete group of isometries (allowing $H^3/\Gamma$ to be compact), these $d=3$
field theories are $N=2$ SCFTs and are dual to solutions of $D=11$ supergravity of the form $AdS_4\times H^3/\Gamma\times S^4$, with the
$S^4$ appropriately fibred over $H^3/\Gamma$. This was shown in
\cite{Gauntlett:2000ng} where the dual holographic solutions,
including the flow across dimensions, were constructed.
Note that when $\Sigma=S^3/\Gamma$, analogous holographic solutions describing the flow across
dimensions were also constructed in \cite{Gauntlett:2000ng}. However, for this case there is not an analogous
$AdS_4\times S^3\times S^4$ solution in the IR (instead one finds a singularity) and hence
the nature of the $d=3$ quantum field theory in the far IR for this case is not clear.

In this paper we will be particularly interested in further studying the $d=3$
SCFTs arising on M5-branes wrapping $H^3/\Gamma$, using holographic techniques.
A primary motivation is that these theories provide a novel arena
for top-down investigations of AdS/CMT. In particular,
we will initiate an investigation of the properties of the $d=3$ $N=2$ SCFTs
at finite temperature and finite charge density (with respect to the abelian $R$-symmetry), by
constructing and analysing appropriate black hole solutions of $D=11$ supergravity.
We show that the high temperature behaviour is described by an (uplifted) AdS-RN type black hole.
One interesting question is whether or not the SCFTs exhibit holographic superconductivity \cite{Gubser:2008px}-\cite{Hartnoll:2008kx}  as found in other top-down supergravity settings using consistent Kaluza-Klein (KK) truncations
\cite{Denef:2009tp}-\cite{Gauntlett:2009bh}.
We will find two instabilities. The first of these involves charged fields implying that there is a new branch
of holographic superconducting black holes with charged hair that spontaneously breaks the $R$-symmetry,
which emerge from the AdS-RN black holes at a branching temperature that
we determine. The second instability only involves neutral fields and implies the existence of
another branch of charged black holes with neutral hair that do not spontaneously break the $R$-symmetry.
Instabilities involving a single neutral scalar field in the background of an
AdS-RN black hole were observed in a bottom up context in
\cite{Hartnoll:2008kx} and arose because the scalar field has a mass that violates the $AdS_2$ BF bound but not the
$AdS_4$ BF bound. In our case the situation is more complicated involving two neutral fields and a massive
vector field and the instability depends on the detailed couplings including the couplings
to the background abelian two-form field strength. This latter feature indicates that
there is some similarity with the bottom-up charged dilaton black holes
studied in \cite{Goldstein:2009cv}.
We will show that the branching temperature for our new charged black holes with neutral scalar and
massive vector hair
is greater than that of the superconducting black holes.

Our results are therefore suggestive that as one cools the $N=2$ SCFT at finite charge density the system will undergo a phase transition, moving to
a phase described by the new charged black holes with neutral hair. However, there are two important caveats.
Firstly, as usual, there could be additional branches of black holes, either inside or outside\footnote{As an example, additional
instabilities appearing outside of the $D=4$ KK truncation
used to construct holographic superconductors (for the special case of the seven-sphere) in
\cite{Gauntlett:2009dn}\cite{Gauntlett:2009bh} were studied in \cite{Bobev:2010ib}. It will be interesting to determine the
implications of these instabilities, as well as those found in
\cite{Denef:2009tp}, for the phase structure for this case.}
the $D=4$ consistent KK truncation that give rise to a phase transition at even higher temperature. Secondly, the conclusion depends on the order of the two phase transitions since
if a phase transition is first order then the critical temperature can be higher
than the branching temperature. In the present context, it is therefore possible that the superconducting black hole
transition is first order and the system moves, discontinuously, from the AdS-RN branch to the superconducting
branch at a higher temperature than the critical temperature associated with the charged
black holes with neutral hair. We will leave a resolution of this interesting issue to future work.

As with several other top down studies of AdS/CMT we will carry out these investigations using
(new) consistent Kaluza-Klein truncations of $D=11$ supergravity. Recall that such truncations have the key property that
the truncated dimensionally reduced theory does not source any of the discarded modes
and hence any solution of the reduced theory uplifts to an exact solution of the higher-dimensional theory.
There has been significant progress in understanding these truncations over the past few years.
For example, it is known that starting with the most general class of supersymmetric $AdS_5$ solutions of Type IIB or $D=11$ supergravity
there are consistent reductions on the internal manifolds to minimal $N=2$ $D=5$ gauged supergravity
\cite{Buchel:2006gb}
\cite{Gauntlett:2006ai} \cite{Gauntlett:2007ma}. Similarly, it has been shown for very
general classes of supersymmetric $AdS_4$ solutions (but not yet the most general) that
there are analogous reductions to minimal $N=2$ $D=4$ gauged supergravity \cite{Gauntlett:2007ma}.
Building on the work of \cite{Maldacena:2008wh}, for the special case of reductions of $D=11$ supergravity on seven-dimensional
Sasaki-Einstein spaces ($SE_7$), it has been shown that the reductions can be extended to include modes that fill out the bosonic part
of an $N=2$ gauged supergravity coupled to a vector multiplet and a hypermultiplet  \cite{Gauntlett:2009zw}. Similar results have also been obtained for
reductions of Type IIB on $SE_5$ \cite{Cassani:2010uw}\cite{Gauntlett:2010vu}\cite{Liu:2010sa} where an interesting enhancement of supersymmetey from
$N=2$ to $N=4$ was observed (see also \cite{Skenderis:2010vz}).
Another development is the addition of the quadratic fermionic sectors to these universal SE truncations \cite{Bah:2010yt}\cite{Bah:2010cu}.
More general truncations for the special case that  $SE_5=T^{1,1}$ have been made in
\cite{Herzog:2009gd}\cite{Cassani:2010na}\cite{Bena:2010pr} and recent results for
$S^5$ and $S^7$ have been obtained in  \cite{Bobev:2010de}\cite{Liu:2010ya} and \cite{Bobev:2010ib}, respectively.

Here we will construct new consistent KK truncations of $D=11$ supergravity on $\Sigma_3\times S^4$ where
$\Sigma_3=H^3, S^3$ or $R^3$ (or a quotient thereof). We will do this in two steps, generalising the work of \cite{Gauntlett:2002rv}.
We first use the well known consistent truncation of $D=11$ supergravity on $S^4$ to
obtain  maximal $SO(5)$ gauged supergravity in $D=7$ \cite{Nastase:1999cb}\cite{Nastase:1999kf}.
We then reduce this $D=7$ gauged supergravity on the above $\Sigma_3$
to obtain $D=4$ $N=2$ gauged supergravities.
More precisely, these KK reductions are at the level of the bosonic fields and we find, for each case,
the bosonic content of an $N=2$ gauged supergravity coupled to a single vector multiplet plus two hypermultiplets.
The scalars in the vector multiplet parametrise the special K\"ahler manifold $SU(1,1)/U(1)$, while the
scalars in the hypermultiplets parametrise the quaternionic K\"ahler space $G_{2(2)}/SO(4)$. We find that
the gauging of the $N=2$ supersymmetry is only in the hypermultiplet sector and we find that
a $U(1)\times\mathbb{R}\subset G_{2(2)}$ is gauged.

We will also use the new consistent truncations to investigate another interesting issue in AdS/CMT:
top down solutions of $D=11$ supergravity that are dual to field theories with Lifshitz symmetry. In \cite{Kachru:2008yh}
a class
of $d+1$-dimensional metrics of the form
\be
ds^2=-r^{2z}dt^2+r^2dx^i dx^i +\frac{dr^2}{r^2},\qquad i=1,\dots d-1
\ee
were proposed to be holographically dual to $d$-dimensional field theories with anisotropic
Lifshitz scaling and dynamical
exponent $z$. It was also shown in \cite{Kachru:2008yh} that these Lif$_{d+1}(z)$ solutions arise as solutions of a
bottom up $d+1$-dimensional phenomenological theory of gravity (see also \cite{Koroteev:2007yp}).
Somewhat surprisingly, it has been
very difficult to embed these solutions into string/M-theory. However, Lif$_4(z=2)$ solutions
of type IIB and Lif$_3(z=2)$ solutions of $D=11$ supergravity were recently constructed in \cite{Balasubramanian:2010uk}
and these were significantly extended in \cite{Donos:2010tu} where supersymmetric solutions
were also presented (which should be stable). Here, using our new consistent truncations,
we will construct\footnote{Note that this solution was constructed prior to those
in \cite{Donos:2010tu} and was announced by one of us (JPG)
at the Non-Perturbative Techniques in Field Theory Symposium in Durham, July 2010.
Just prior to the submission of this paper, other constructions of Lifshitz solutions were
presented in \cite{Gregory:2010gx}.}
a new Lif$_4(z)\times H^3/\Gamma\times S^4$ solution of $D=11$ supergravity
with dynamical exponent $z=39.05...$ (determined by solving some algebraic equations numerically).

The plan of the rest of the paper is as follows. In section 2, we first briefly review the consistent
KK truncation of $D=11$ supergravity on $S^4$ to $D=7$ $SO(5)$ gauged supergavity \cite{Nastase:1999cb}\cite{Nastase:1999kf} using the presentation of \cite{Cvetic:2000ah}. In section 3 we construct the consistent
truncation of $D=7$ $SO(5)$ gauged supergravity on $H^3,S^3,R^3$ (or a quotient by $\Gamma$ thereof).
In section 4 we show that
the $D=4$ truncated theory is the bosonic part of
an $N=2$ gauged supergravity, coupled to a vector multiplet and two hypermultiplets
and we elucidate the gauging. The natural degrees of freedom required to exhibit the $N=2$
supersymmetry in section 4 require some dualisation of the degrees of freedom that naturally appear in
the uplifting formulae given in section 3. We emphasise that apart from section 4, we use the variables given
in section 3.

In section 5 we recall the supersymmetric $AdS_4$ solution
of the $D=4$ reduced theory (for the case of $H^3/\Gamma$)
which uplifts to the $AdS_4\times H^3/\Gamma\times S^4$ solution
dual to the $N=2$ SCFT on the wrapped M5-branes. Within the $D=4$ theory we analyse
the linearised spectrum of fluctuations and show how they correspond to $OSp(2|4)$ multiplets
of operators in the dual SCFT. The reduced $D=4$ theory has another non-supersymmetric $AdS_4$ vacuum
which uplifts to a non-supersymmetric $AdS_4\times H^3/\Gamma\times S^4$ solution
of $D=11$ supergravity first found in \cite{Gauntlett:2002rv}. For this solution we also analyse the mass
spectrum and find that within
the $D=4$ truncation there are no unstable modes. Section 6 briefly considers some
additional truncations of the $D=4$ theory.

In section 7 we switch gears and study the $N=2$ SCFT, dual to the supersymmetric
$AdS_4\times H^3/\Gamma\times S^4$ solution, at finite temperature and finite chemical potential.
At high temperatures the system is described by an uplifted AdS-RN type black hole with flat spatial
horizon (often called a black brane). We show that
at zero temperature there are two kinds of instabilities, one of which involves charged fields
and is associated with holographic superconductivity and the other just involves neutral fields.
By studying the fluctuations about
the AdS-RN black hole at finite temperature we then deduce the
temperatures at which the two new branches of black holes appear, finding that
the non-superconducting charged black holes with neutral hair have a higher branching temperature than
the superconducting black holes.

In section 8 we construct the Lif$_4(z\sim 39)\times H^3/\Gamma\times S^4$ solution of
$D=11$ supergravity. We conclude with some discussion in section 9. The paper contains two
appendices: in appendix A we have presented some details of the consistent KK truncation,
including the full set of $D=4$ equations of motion, given in \eqref{firstsetone}-\eqref{4dEinstein},
that are used throughout this paper.
In appendix B we have made some comments concerning an unconventional presentation of a
massive vector field that emerges in our truncation. In appendix C we
have recorded some details on the coset $G_{2(2)}/SO(4)$ which we use in elucidating the $N=2$
supersymmetry of the reduced $D=4$ theory.

\section{Maximal $D=7$ gauged supergravity and uplift to $D=11$ supergravity on $S^4$}

The bosonic fields of $D=7$ gauged supergravity \cite{Pernici:1984xx} consist of a metric, $g_7$, $SO(5)$
Yang-Mills fields $A^{ij}$, $i,j=1,\dots 5$, five three-forms $S_{(3)}^i$ transforming in the  ${\bf 5}$
of $SO(5)$ and fourteen scalar fields, given by the symmetric unimodular matrix
$T_{ij}$, which parametrise the coset $SL(5,\mathbb{R})/SO(5)$.
The seven-form Lagrangian for the bosonic fields is given by
\bea
{\cal L}_7 &=& R\, {*\oneone} -
\ft14 T^{-1}_{ij}\, {*D T_{jk}}\wedge
T^{-1}_{k\ell}\, D T_{\ell i}
-\ft1{4}\, T^{-1}_{ik}\, T^{-1}_{j\ell}\, {* F_\2^{ij}}\wedge F_\2^{k\ell}
-\ft12 T_{ij}\, {*S_\3^i}\wedge S_\3^j \nn
&&+ \ft{1}{2g} S_\3^i\wedge DS_\3^i -
\ft{1}{8g}  \ep_{i j_1\cdots j_4}\, S_\3^i\wedge F_\2^{j_1 j_2}\wedge
F_\2^{j_3 j_4} +
\ft{1}g \Omega_\7 - V\, {*\oneone}\,,\label{d7lag}
\eea
where
\bea
DT_{ij}& \equiv& dT_{ij} + g A_\1^{ik}\, T_{kj} + g A_\1^{jk}\, T_{ik}\nn
D S_\3^i &\equiv &dS_\3^i + g\, A_\1^{ij}\wedge S_\3^j\nn
F_\2^{ij} &\equiv& dA_\1^{ij} + g A_\1^{ik}\wedge A_\1^{kj}\,,
\eea
the potential $V$ is
given by
\be
V = \ft12  g^2 \Big(2 T_{ij}\, T_{ij} - (T_{ii})^2 \Big)\,,
\ee
and $\Omega_\7$ is a Chern-Simons type of term built from the
Yang-Mills fields, which has the property that its variation with
respect to $A_\1^{ij}$ gives
\be
\delta \Omega_\7 =
\ft34 \delta_{i_1 i_2 k\ell}^{j_1 j_2 j_3 j_4}\, F_\2^{i_1 i_2}\wedge
F_\2^{j_1 j_2}\wedge  F_\2^{j_3 j_4}\wedge \delta A_\1^{k\ell}\,.
\ee
An explicit expression can be found in \cite{Pernici:1984xx}.

Any solution to the associated $D=7$ equations of motion, which are given in
appendix A, gives rise to a solution of $D=11$ supergravity
\cite{Nastase:1999cb}\cite{Nastase:1999kf}. Using the notation of
\cite{Cvetic:2000ah}, the $D=11$ metric and four-form field strength are given by
\bea
d s_{11}^2 &=& \Delta^{1/3}\, ds_{7}^2 + \fr1{g^2}\Delta^{-2/3}\,
T^{-1}_{ij}\, D\mu^i\, D\mu^j\,,\label{metel}
\eea
\bea
&&G_\4 = \frac{\Delta^{-2}}{g^34!}\, \ep_{i_1\cdots i_5}\, \Big[
-  U\,  \mu^{i_1} D\mu^{i_2}\wedge D\mu^{i_3}\wedge D\mu^{i_4}\wedge
D\mu^{i_5}\nn
&& + 4 \, T^{i_1 m}\, DT^{i_2 n}\, \mu^m\, \mu^n\,
D\mu^{i_3}
\wedge D\mu^{i_4} \wedge D\mu^{i_5}+ 6g \Delta F_\2^{i_1 i_2} \wedge
D\mu^{i_3}\wedge D\mu^{i_4}\, T^{i_5 j}\, \mu^j \Big]\nn
&&- T_{ij}\,
{*S_\3^i}\, \mu^j + \fft1{g}\, S_\3^i \wedge D\mu^i\,,\label{4form}
\eea
where $\mu^i$, $i= 1, \ldots, 5$ are constrained coordinates on $S^4$ satisfying $\mu^i\ \mu^i \equiv 1$, and
\bea
U \equiv 2 T_{ij}\, T_{jk}\, \mu^i\, \mu^k - \Delta\, T_{ii}\,, \qquad
\Delta \equiv T_{ij}\, \mu^i\, \mu^j\,,\qquad
D\mu^i \equiv d\mu^i + g A_\1^{ij}\, \mu^j\, \,.
\eea

For example, the basic $AdS_7$ vacuum solution of $D=7$ supergravity, with $A^{ij}_\1=S^i_\3=0$ and $T_{ij}=\delta_{ij}$ uplifts to the maximally supersymmetric $AdS_7\times S^4$ solution. Of more interest to this paper
is the supersymmetric $AdS_4\times H^3$ solution found in
\cite{Gauntlett:2000ng}.
This solution uplifts to an $AdS_4\times H^3\times S^4$ solution, with a warped product metric and
the $S^4$ non-trivially fibred over the $H^3$ factor. The solution preserves eight supercharges and
the $H^3$ factor can be replaced with an arbitrary quotient $H^3/\Gamma$, possibly compact, and still
preserve all supersymmetry. When $H^3/\Gamma$ is compact
these solutions are dual to $N=2$ superconformal
field theories in three spacetime dimensions that arise on the non-compact part of fivebranes
wrapping special Lagrangian three-cycles
$H^3/\Gamma$.
We will recall this solution in section 5 below.

It is worth pointing out that the conventions for $D=7$ gauged supergravity used in
\cite{Cvetic:2000ah} and in this paper, slightly differ from those used in \cite{Pernici:1984xx},
which were also used in \cite{Gauntlett:2000ng}. In particular $g^{here}=m^{there}$ (and one should
be careful since $g^{there}=2m^{there}$) and also $A^{here}=2A^{there}$.

\section{Consistent KK truncation of $D=7$ gauged supergravity on $S^3$, $H^3$ or $R^3$}
\label{conkktrunc}


We now construct the consistent KK ansatz for the reduction of $D=7$ supergravity on
$\Sigma_3=S^3, H^3$ or $R^3$ (or a quotient thereof), generalising that of \cite{Gauntlett:2002rv}.
For the $D=7$ metric we take
\be \label{KKmetricf}
ds^2_7 =e^{-6\phi}ds^2_4+e^{4\phi}ds^2(\Sigma_3)
\ee
where $ds^2_4$ is an arbitrary metric on the $D=4$ external spacetime,
$ds^2(\Sigma_3)$ is the maximally symmetric metric on $S^3$ or $H^3$ or $T^3$ (or a quotient thereof)
normalised so that
the Ricci tensor is $lg^2$ times the metric, for $l=+1,-1$ or $0$ respectively, and $\phi$ is a real ``breathing mode" scalar field
defined on the $D=4$ external spacetime.

To construct the ansatz for the remaining fields we introduce
an orthonormal frame, $\bar e^a$, for $ds^2(\Sigma_3)$, and let
$\bar{\omega}^{ab}$ be the corresponding Levi-Civita spin connection:
\bea
d\bar e^a+\bar{\omega}^a{}_b\wedge \bar e^b=0\,.
\eea
Then for the $D=7$ $SO(5)$ vector fields, we consider an $SO(3)\times SO(2)$
split, with $a,b=1,2,3$ and $\alpha,\beta=4,5$ and set
\bea \label{KKvectors}
&& A^{ab}_\1 = \tfrac{1}{g}\bar{\omega}^{ab} +
\beta\epsilon_{abc} \ \bar{e}^c
\nn
&& A^{a\alpha}_\1 = -A^{\alpha a}_\1 =   \theta^{\alpha} \bar{e}^a
\nn
&& A^{\alpha \beta}_\1 = \epsilon^{\alpha \beta}  A_1\,.
\eea
This ansatz incorporates a
scalar field $\beta$, two scalar fields $\theta^\alpha$ and a vector field $A_1$, all defined on the
$D=4$ external spacetime.
We note that we take $\epsilon_{45} =-\epsilon_{54} = 1$. Indices $a$ and $\alpha$ are raised and lowered with $\delta_{ab}$ and $\delta_{\alpha \beta}$, respectively.

The split of $SO(5)$ into $SO(3)\times SO(2)$ is also used in the
ansatz for the five $D=7$ three-forms and the fourteen $D=7$ scalars. Specifically,
for the 3-form fields $S^i_\3$ we take:
\bea \label{KK3formf}
S^a_\3 &=& B_2 \wedge \bar{e}^a  + C_1 \wedge \epsilon_{abc} \bar{e}^b \wedge \bar{e}^c
\nn
S^\alpha_\3 &=& h_3^\alpha +  g \chi^\alpha \mathrm{vol}(\Sigma_3)\,.
\eea
where $B_2, C_1,\chi^\alpha,h^\alpha_3$ are 2-,1-,0-, and 3-forms in $D=4$, respectively.
For the scalars $T_{ij}$ parametrising the coset $SL(5,\mathbb{R})/SO(5)$ we choose:
\begin{equation}\label{KKscalarsf}
T_{ab}= e^{-4\lambda} \delta_{ab} \; , \quad T_{a\alpha} =0 \; , \quad T_{\alpha \beta} =e^{6\lambda} \cT_{\alpha \beta}\,,
\end{equation}
where $\lambda$ is a scalar and the symmetric, unimodular matrix $\cT_{\alpha \beta}$, parametrises
the coset $SL(2,\mathbb{R})/SO(2)$ (and thus contains two scalar degrees of freedom), all in $D=4$.

In summary, the above ansatz incorporates the following $D=4$ content:
9 scalars $\phi,\lambda, \cT_{\alpha \beta}, \beta,\theta_\alpha,\chi_\alpha$,
two one-forms $A_1, C_1$, one two-form $B_2$, two three-forms $h^\alpha_3$, plus the metric.
 As we will see in the next section, after some field redefinitions,
 these arrange themselves into bosonic fields of the following
$N=2$ multiplets: a gravity multiplet (metric plus, a vector), a vector multiptlet (vector plus two scalars)
and two hypermultiplets (eight scalars).

We next substitute our ansatz into the $D=7$ equations of motion. After some arduous calculation
we find that they are equivalent to unconstrained equations of motion for the $D=4$ fields,
thus demonstrating the consistency of the ansatz. We have presented a few details in appendix A, and
the equations of motion are given in \eqref{firstsetone}-\eqref{4dEinstein}.
We have also verified that these equations of motion can all be derived from
the $D=4$ (four-form) Lagrangian given by
\begin{eqnarray} \label{4dLagrangian}
2 {\cal L} =  {\cal L}_\textrm{kin}  +  {\cal L} _\textrm{pot}+{\cal L}_\textrm{top}\,,
\end{eqnarray}
where
%
\begin{eqnarray} \label{4dLagKin}
{\cal L}_\textrm{kin} &=& R^{(4)} \textrm{vol}_4 + 30 d \phi \wedge *d\phi + 30 d \lambda \wedge *d\lambda + \ft14 \Tr(\cT^{-1}D\cT \wedge * \cT^{-1}D\cT)  \nonumber \\
&& +\tfrac{3}{2}e^{8\lambda-4\phi}d \beta \wedge *d\beta
+\tfrac{3}{2}e^{-2\lambda-4\phi} D \theta^T \wedge * \cT^{-1} D\theta +\tfrac{1}{2}e^{6\lambda+12\phi} h^T_3 \wedge * \cT h_3
  \nonumber \\
&& -\tfrac{1}{2}e^{-12\lambda+6\phi}F_2 \wedge * F_2 -\tfrac{3}{2}e^{-4\lambda+2\phi}B_2 \wedge *B_2 +6 e^{-4\lambda-8\phi}C_1 \wedge *C_1\,,
\end{eqnarray}
\begin{eqnarray} \label{4dLagPot}
{\cal L}_\textrm{pot} &=& g^2  \Big\{ 3 l e^{-10\phi}  -\ft{3}{8} e^{8\lambda-14\phi} (l -2 \beta^2 - 2 \theta^T \theta)^2 \nonumber \\
&& \qquad   +\ft12  e^{-6\phi} \left[ 3e^{-8\lambda}+e^{12\lambda}[ (\Tr\cT)^2-2\Tr(\cT\cT)]+6e^{2\lambda}\Tr\cT \right]  \nonumber \\
&& \qquad  -\tfrac{3}{2} e^{-10\phi}\left[e^{10\lambda}(\theta^T\cT\theta)-2\theta^T\theta+e^{-10\lambda}
(\theta^T\cT^{-1}\theta)\right] \nonumber \\
&& \qquad  -6e^{-2\lambda-14\phi} \beta^2 (\theta^T\cT^{-1}\theta) -  \ft{1}{2} e^{6\lambda-18\phi}(\chi^T\cT\chi) \Big\} \textrm{vol}_4
\end{eqnarray}
and
\bea\label{topl}
g\,{\cal L}_\textrm{top}&=& 6\,C_{1}\wedge\left(dB_{2}-g\theta^{T}h_{3} \right)-6\, C_{1}\wedge d\beta\wedge F_{2} -3g\beta\,B_{2}\wedge B_{2} \nonumber\\
&&-3\,B_{2}\wedge D\theta^{T}\wedge\varepsilon D\theta
+\, g \chi^{T}Dh_{3}-6g\beta\, h_{3}^{T}\varepsilon\theta\wedge d\beta\nn
&&+2g\,\beta^{3}F_{2}\wedge F_{2}
+3\,\beta D\theta^{T}\wedge\varepsilon D\theta\wedge F_{2}\nn
&&+\frac{g}{2}\left(l-2\beta^{2}-2\,\theta^{T}\theta \right)\left[-3\,B_{2}\wedge F_{2}
+3\,h_{3}^{T}\wedge\varepsilon D\theta
 +3\,\beta F_{2}\wedge F_{2}\right]\,.
\eea
In these expressions we have defined
\be F_2=dA_1
\ee
and we have introduced the $SO(2)$--covariant differential $D$ which
acts on  $SO(2)$ doublets, $X^\alpha$, as
 \be\label{covddoublet}
 DX^\alpha=dX^\alpha+g\epsilon_{\alpha\beta}A_1\wedge X^\beta
 \ee
 and on the coset scalars as
 \be
 D \cT^{\alpha\beta} = d \cT^{\alpha\beta} +g\epsilon_{\alpha\gamma} A_1  \cT^{\gamma\beta} + g\epsilon_{\beta\gamma} A_1  \cT^{\alpha\gamma}\,.
 \ee
The $D=4$ Lagrangian (\ref{4dLagrangian}) is locally $SO(2)$ invariant.

We note that upon setting $\theta=\beta=\chi=0$ and $\cT_{\alpha\beta}=\delta_{\alpha\beta}$ we obtain the consistent truncation
studied in \cite{Gauntlett:2002rv}. In particular, $ {\cal L}$ agrees with (2.12), (2.13) of \cite{Gauntlett:2002rv}. Additional consistent KK truncations will be considered in section \ref{addcontrunc}.

In the next section we will demonstrate that this $D=4$ reduced theory comprises the bosonic sector of an $N=2$ $D=4$ gauged supergravity coupled to a vector multiplet and two hypermultiplets. The factor of 2 appearing in \eqref{4dLagrangian} is incorporated to facilitate comparison with some standard conventions used in the $N=2$ literature. We would like to emphasise that to do so will require some field redefinitions. In particular, we will see that,
essentially, $B_2$ and $C_1$, which we observe do not have the usual kinetic energy terms
in \eqref{4dLagrangian}, will be replaced by a vector and a scalar, and the auxiliary three-forms $h_3^\alpha$
will be eliminated. We emphasise that in later sections when we construct new solutions of the $D=4$ theory
we will work in the variables given in \eqref{4dLagrangian}-\eqref{topl} (i.e. we will
analyse the equations of motion \eqref{firstsetone}-\eqref{4dEinstein}) as these variables
are the easiest ones to uplift to to $D=11$. It is straightforward to translate to the standard $N=2$ variables given
in the next section. We have made some clarifying comments on the $B_2$, $C_1$ system in a simplified setting in
appendix \ref{lastapp}.

\section{Explicit $N=2$ supersymmetry} \label{sec:susy}

\subsection{New variables}
In order to display the $N=2$ supersymmetry we will need to change to new variables (see also appendix \ref{lastapp}). To begin with, we first
perform the trivial relabelling
\be
B_2 \rightarrow H_2,\qquad C_1 \rightarrow G_1,\qquad h_3^\alpha \rightarrow F_3^\alpha
\ee
with the capital letters $H$ and $F$ used to indicate that the objects are, or will become, field strengths, or dual field strengths in the new variables.

We continue by defining a new two-form field strength defined by
\begin{equation} \label{tildeH2def}
 \tilde H_2 \equiv e^{-4\lambda+2\phi}*H_2+2\beta H_2 - \beta^2 F_2 \; .
\end{equation}
Some manipulation of equations (\ref{firstsetone}), (\ref{firstsettwo}), (\ref{secondsetone}), (\ref{secondsettwo})
allows us to write the following Bianchi identities
\begin{eqnarray} \label{Bianchisfullg}
&& d \tilde H_2 =0 \; , \nn
&& dG_1 -\tfrac12 g \tilde H_2 -\tfrac14 g (l-2\theta^T \theta) F_2 - \tfrac12 \epsilon_{\alpha \beta} D\theta^\alpha \wedge D \theta^\beta =0\,.
\end{eqnarray}
%
These can be integrated to give
\begin{eqnarray} \label{H2tildesolvfg}
&& \tilde H_2 =d \tilde B_1 \; , \\
&&  G_1 = D \ab + \tfrac12 \epsilon_{\alpha \beta}  \theta^\alpha D\theta^\beta \label{G1solvfg}\,,
\end{eqnarray}
where we have defined
\begin{equation} \label{covDb}
D \ab \equiv d \ab + \tfrac14 gl A_1 + \tfrac12 g \tilde B_1\,.
\end{equation}
%

Next, notice that equation (\ref{firstsetfour}) can be written as
\begin{eqnarray} \label{starF3fg}
&& e^{6\lambda+12\phi}  *(\cT F_3)_\alpha =- D \left[\chi_\alpha +  \tfrac{3}{2}  \epsilon_{\alpha \beta} \theta^\beta (l-2\beta^2-2\theta^T\theta) \right]-6 \theta_\alpha G_1 - 6 \epsilon_{\alpha \beta} \theta^\beta \theta^\gamma D \theta_\gamma \; .
\nonumber \\ &&
\end{eqnarray}
Introducing a scalar
\begin{equation} \label{hfg}
\xi_\alpha \equiv - \chi_\alpha -   \epsilon_{\alpha \beta} \theta^\beta (\tfrac{3}{2} l-3\beta^2 -\theta^T\theta)\,,
\end{equation}
with one-form field strength
\begin{equation} \label{F1fg}
F_1^\alpha \equiv D\xi^\alpha -6 \theta^\alpha D \ab - \epsilon_{\gamma
\delta}\theta^\alpha \theta^\gamma D\theta^\delta  \; ,
\end{equation}
one can use (\ref{G1solvfg}) to show that (\ref{starF3fg}) is equivalent to
\begin{equation} \label{F3fg}
F_3 = e^{-6\lambda-12\phi} \cT^{-1} * F_1 \; .
\end{equation}

In terms of the new variables, we can identify the degrees of freedom of our theory as a metric, two vectors $A_1, \tilde B_1$, and ten scalars $\phi,\lambda, \cT_{\alpha \beta}, \beta,\theta_\alpha, \ab , \xi_\alpha$. The $D=4$ field equations (\ref{firstsetone})--(\ref{4dEinstein}) can be translated to the new variables by using equations (\ref{hfg}), (\ref{F1fg}), (\ref{F3fg}) and solving for $H_2$ from (\ref{tildeH2def}) as
\begin{equation}
H_2=\frac{1}{4\beta^2+e^{-8\lambda+4\phi}}\left[2\beta (\tilde H_2+ \beta^2F_2)-e^{-4\lambda+2\phi}*(\tilde H_2+\beta^2F_2)\right] \; .
\end{equation}
In order to write down the Lagrangian, it also proves convenient to define the two dilatons
\begin{eqnarray}
\varphi_0 = -4\lambda +2\phi \; , \qquad \quad \varphi_1 = 2\sqrt{3}(
\lambda +2\phi) \; ,
\end{eqnarray}
and the axion-dilaton
\begin{equation} \label{taudef}
\tau = \beta + i e^{\varphi_0} \; .
\end{equation}
The Lagrangian that gives rise to the equations of motion (\ref{firstsetone}), (\ref{firstsetthree}), (\ref{secondsettwo}), (\ref{betaeom})--(\ref{4dEinstein}) upon variation of $\tilde B_1$, $\xi^\alpha$, $\ab$, $\beta$, $\theta^\alpha$, $A_1$, $\lambda$, $\cT_{\alpha \beta}$, $\phi$ and the metric, respectively, is
\begin{equation} \label{lagfgt}
{\cal L} = \tfrac{1}{2} R^{(4)} \textrm{vol}_4 + {\cal
L}_{\textrm{VM}} + {\cal L}_{\textrm{HM}} +  {\cal L}_{\textrm{pot}} \; ,
\end{equation}
with
\begin{equation} \label{lagVMfgt}
\begin{aligned}
  {\cal L}_{\textrm{VM}} &= \tfrac34 d \varphi_0 \wedge *d \varphi_0
+\tfrac{3}{4}e^{-2\varphi_0}d \beta \wedge *d\beta
      \\ & \quad
      + \tfrac{3}{4} \textrm{Im} \left[ (\tau+\beta)^{-1} \right]
         \big( \tilde{H}_2+\beta^2F_2\big)
         \wedge\ast\big(\tilde{H}_2+\beta^2F_2\big)
      \\ & \quad
      + \tfrac{3}{4}  \textrm{Re} \left[ (\tau+\beta)^{-1} \right]
      \big(\tilde{H}_2+\beta^2F_2\big)\wedge\big(\tilde{H}_2+\beta^2F_2\big)
      \\ & \quad
      -\tfrac{1}{4} e^{3 \varphi_0} F_2\wedge\ast F_2
      - \tfrac{3}{2}\beta\tilde{H}_2\wedge F_2-  \tfrac{1}{2} \beta^3 F_2\wedge F_2 \; ,
\end{aligned}
\end{equation}
\begin{eqnarray} \label{lagHMfgt}
{\cal L}_{\textrm{HM}} &=& \tfrac14 d \varphi_1 \wedge *d \varphi_1 +
\ft18 \Tr(\cT^{-1}D\cT \wedge * \cT^{-1}D\cT)
+\tfrac{3}{4}e^{-\frac{1}{\sqrt{3}} \varphi_1} D \theta^T \wedge * \cT^{-1} D\theta  \nonumber \\
&& +\tfrac{1}{4}e^{-\sqrt{3} \varphi_1} F^T_1 \wedge * \cT^{-1} F_1
  +3 e^{-\frac{2}{\sqrt{3}} \varphi_1}G_1 \wedge *G_1\,,
\end{eqnarray}
where $G_1$, $F_1$ are given in (\ref{G1solvfg}),  (\ref{F1fg}) and
\begin{eqnarray} \label{lagPotfgt}
{\cal L}_\textrm{pot} &=& \tfrac{1}{2} g^2  \Big\{ 3 l e^{-10\phi}  -\ft{3}{8} e^{8\lambda-14\phi} (l -2 \beta^2 - 2 \theta^T \theta)^2 \nonumber \\
&& \qquad   +\ft12  e^{-6\phi} \left[ 3e^{-8\lambda}+e^{12\lambda}[ (\Tr\cT)^2-2\Tr(\cT\cT)]+6e^{2\lambda}\Tr\cT \right]  \nonumber \\
&& \qquad  -\tfrac{3}{2} e^{-10\phi}\left[e^{10\lambda}(\theta^T\cT\theta)-2\theta^T\theta+e^{-10\lambda}
(\theta^T\cT^{-1}\theta)\right] \nonumber \\
&& \qquad  -6e^{-2\lambda-14\phi} \beta^2 (\theta^T\cT^{-1}\theta) -  \ft{1}{2} e^{6\lambda-18\phi}(\chi^T\cT\chi) \Big\} \textrm{vol}_4\,,
\end{eqnarray}
where (\ref{hfg}) should be used to write $\chi^\alpha$ in terms of  $\xi^\alpha$, $\theta^\alpha$ and $\beta$.

The $D=4$ Lagrangian (\ref{lagfgt}) has now a local $U(1) \times
\mathbb{R}$  symmetry. Note, in particular, from (\ref{covDb}) that a
non-compact, local $\mathbb{R}$ shift of the scalar field $a$ can be cancelled
by a gauge transformation of the vector $l A_1 + 2 \tilde B_1$. Furthermore, the
scalar potential (\ref{lagPotfgt}) does not depend on $a$.
Note that the Lagrangian (\ref{4dLagrangian}), with its local $SO(2)\cong U(1)$ symmetry,
corresponds, in part, to an  $\mathbb{R}$-gauge-fixed version of (\ref{lagfgt}).

\subsection{$N=2$ supersymmetry}

We will now show that our $D=4$ reduced theory corresponds to the bosonic part of
$D=4$ $N=2$ supergravity coupled to a vector multiplet and two hypermultiplets, with an Abelian $  U(1) \times \mathbb{R}$ gauging in the hypermultiplet sector. The scalar manifold is the symmetric space
\begin{equation}
\frac{SU(1,1)}{U(1)} \times \frac{G_{2(2)}}{SO(4)} \; ,
\end{equation}
where the first factor is the special K\"ahler manifold parametrised by the two real scalars in the vector multiplet, and the second factor is the quaternionic-K\"ahler manifold parametrised by the eight real scalars in the hypermultiplets.

We will show this by casting the Lagrangian (\ref{lagfgt})--(\ref{lagPotfgt}) into the canonical $N=2$ form
(see, for example, \cite{Andrianopoli:1996vr}\cite{Andrianopoli:1996cm})
\begin{eqnarray} \label{4dLagsplit}
{\cal L} &=& \tfrac{1}{2} R^{(4)} \textrm{vol}_4 + {\cal L}_{\textrm{VM}} + {\cal L}_{\textrm{HM}} -V \textrm{vol}_4\,,
\end{eqnarray}
where
\begin{equation} \label{LagVMcan}
{\cal L}_{\textrm{VM}} = g_{\tau \bar \tau}d \tau \wedge \ast d\bar\tau
        + \tfrac{1}{2}\textrm{Im}
\mathcal{N}_{IJ}F^I\wedge\ast F^J
      + \tfrac{1}{2}\textrm{Re} \mathcal{N}_{IJ}F^I\wedge F^J
\end{equation}
is the piece corresponding to the (ungauged) vector multiplet,
\begin{equation} \label{LagHMcan}
{\cal L}_{\textrm{HM}} = h_{uv} Dq^u\wedge\ast Dq^v
\end{equation}
the piece corresponding to the gauged hypermultiplets, and $V$ is the scalar potential. We now elaborate on
each of these terms.

In (\ref{LagVMcan}), $\tau$ is a complex coordinate and $g_{\tau \bar \tau}$ a K\"ahler metric, with K\"ahler potential $K_V$, on the special K\"ahler manifold $SU(1,1)/U(1)$; $F_I=dA^I$, $I=0,1$ are the abelian two-form
field strengths of the graviphoton and vector in the vector multiplet; and $\mathcal{N}_{IJ}$ is a $\tau$-dependent matrix, specified by supersymmetry, governing the couplings of  the scalars to the gauge kinetic terms and the Chern-Simons terms. Specifically, if $X^I$ are homogeneous coordinates on the special K\"ahler manifold and a prepotential
$\mathcal{F}$ exists, in terms of which the K\"ahler potential can be written as
\begin{equation} \label{Kpotfromprepot}
   K_V = -\log\left(i\bar{X}^I \mathcal{F}_I
       -i X^I\bar{\mathcal{F}}_I \right) \ ,
\end{equation}
then $\mathcal{N}_{IJ}$ is given by
\begin{eqnarray} \label{NIJgaugedDef}
   \mathcal{N}_{IJ} \equiv \ \bar{\mathcal{F}}_{IJ}
       + 2i\frac{(\textrm{Im}\mathcal{F}_{IK})(\textrm{Im}\mathcal{F}_{JL})X^K X^L}
       {(\textrm{Im}\mathcal{F}_{AB})X^A X^B}\,,
\end{eqnarray}
where $\mathcal{F}_I=\partial_I \mathcal{F}$ and
$\mathcal{F}_{IJ}=\partial_I\partial_J\mathcal{F}$ are the derivatives of the prepotential with respect to $X^I$.

In (\ref{LagHMcan}), $q^u$, $u=1, \ldots, 8$, are coordinates and $h_{uv}$ the homogeneous
metric on the quaternionic-K\"ahler manifold $G_{2(2)}/SO(4)$,
normalised so that its Ricci tensor is $-2(2+n_H)=-8$ times the metric, where $n_H=2$ is the number of hypermultiplets. The covariant derivatives of $q^u$ are defined in terms of two specific Killing vectors $k^u_I$, $I=0,1$, of $G_{2(2)}/SO(4)$ as  $Dq^u = d q^u -g k^u_I A^I$.

Finally, when a gauging is turned on in the hypermultiplet sector only, as in the present case, the $N=2$ scalar potential $V$ in \eqref{4dLagsplit} is given by
\begin{equation} \label{scalarpotcan}
V = e^{K_{V}}X^{I}\bar{X}^{J}4h_{uv}k^{u}_{I}k^{v}_{J}
-\left(\tfrac{1}{2}\mathrm{Im}\mathcal{N}^{-1IJ}
+4e^{K_{V}}X^{I}\bar{X}^{J} \right)P^{x}_{I}P^{x}_{J} \; ,
\end{equation}
where the only symbols that remain to be defined are the momentum maps $P^{x}_{I}$.
First recall that the quaternionic-K\"ahler manifold has $Sp(1)\times Sp(n_H)$ holonomy
and the $Sp(1)$ factor is associated to the existence of a triplet of complex structures.
Let $\omega^{x}$ be the $Sp(1)$ part of the spin connection
and $K^x$ the corresponding curvature
(see appendix \ref{hyperMS} for more details). Then for each of the two Killing vectors $k^u_I$, $I=0,1$,
of $G_{2(2)}/SO(4)$ along which the gauging is turned on,
the scalars $P^{x}_{I}$, $x=1,2,3$, are a triplet of potentials for the $Sp(1)$ part of the curvature $K^x$ of $G_{2(2)}/SO(4)$
along $k^u_I$ satisfying\footnote{Note that in \cite{Gauntlett:2009zw} there should be a factor of 2 appearing on the left hand side of
equation (C.7) and a factor of 1/2 on the right hand side of (C.13).}:
\begin{equation} \label{mommapsdef}
2 \imath_{k_{I}}K^{x} = DP_I^x \equiv \left( dP^{x}_{I}+\epsilon^{xyz}\omega^{y}P^{z}_{I} \right) \; .
\end{equation}

We now show that our $D=4$ Lagrangian (\ref{lagfgt})--(\ref{lagPotfgt}) is of this form.

\subsubsection{Vector multiplet}

Let us first deal with ${\cal L}_{\textrm{VM}}$. Introducing the homogeneous coordinates $X^I=(1,\tau^2)$,  $I=0,1$, with $\tau$ defined in (\ref{taudef}), we can write down the holomorphic prepotential
\begin{equation} \label{prepot}
   \mathcal{F} = \sqrt{X^0(X^1)^3} \ .
\end{equation}
From (\ref{Kpotfromprepot}), this gives the K\"ahler potential
\begin{equation}
   K_V = - \log i(\tau-\bar{\tau})^3 + \log 2 \ ,
\end{equation}
from which we obtain the K\"ahler metric
\begin{equation}
g_{\tau \bar \tau} = \partial_{\tau}\partial_{\bar \tau} K_V = -\frac{3}{(\tau - \bar \tau)^2}
\end{equation}
on $SU(1,1)/U(1)$.

Next, equation (\ref{NIJgaugedDef}) allows us to compute
\begin{eqnarray} \label{NIJgauged}
   \mathcal{N}_{IJ}
      = \ \frac{1}{2(\tau+\beta)} \begin{pmatrix}
           -\tau^3\bar\tau & 3 \beta \tau \\
           3 \beta \tau\ & 3
        \end{pmatrix} \; .
\end{eqnarray}
Finally, defining the Abelian gauge fields $A^I=(A_1,-\tilde B_1)$, $I=0,1$, with field strengths $ F^I=dA_1^I=(F_2,-\tilde H_2)$ (see (\ref{H2tildesolvfg})), it is straightforward to now show
that the Lagrangian (\ref{lagVMfgt}) can indeed be cast in the canonical form (\ref{LagVMcan}).

Observe that exactly the same vector multiplet structure arises in the $D=4$, $N=2$ theory obtained from consistent truncation of $D=11$ supergravity on an arbitrary Sasaki-Einstein seven-fold \cite{Gauntlett:2009zw}. Indeed, we find agreement with the analysis of section 2.3 of \cite{Gauntlett:2009zw} with the identifications
$\varphi_0 = 2U+V$, $\beta =h$ and identical $H_2$ and $F_2$. The quantities corresponding to
$X^I$, $\mathcal{F}$, $K_V$ and $\mathcal{N}_{IJ}$ were denoted with tildes in \cite{Gauntlett:2009zw}.

\subsubsection{Gauged hypermultiplets}

We next show that the Lagrangian ${\cal L}_{\textrm{HM}}$ corresponds to a gauged non-linear sigma model with $G_{2(2)}/SO(4)$ target space. Some details about this coset space are given in appendix \ref{hyperMS}.
We find it useful here to use the $SL(2, \mathbb{R})/SO(2)$ Iwasawa parametrisation for $\cT_{\alpha \beta}$,
\be\label{Npar}
\cT=\begin{pmatrix}
            e^{\varphi_2} &\ & \zz e^{\varphi_2}\\
            \zz e^{\varphi_2} &\ & e^{-\varphi_2}+\zz^2 e^{\varphi_2}
        \end{pmatrix} \; ,
\ee
which allows us to write the Lagrangian (\ref{lagHMfgt}) as
\begin{eqnarray} \label{HMpara}
{\cal L}_{\textrm{HM}} &=& \tfrac14 d \varphi_1 \wedge *d \varphi_1
+\tfrac14 D \varphi_2 \wedge *D \varphi_2 + \tfrac14 e^{2\varphi_2} D
\zz \wedge *D \zz
+\tfrac{3}{4}e^{-\frac{1}{\sqrt{3}}\varphi_1-\varphi_2} D \theta^1
\wedge * D\theta^1 \nonumber \\
&&   +\tfrac{3}{4}e^{-\frac{1}{\sqrt{3}}\varphi_1+\varphi_2} (D\theta^2
- \zz D \theta^1) \wedge * (D\theta^2 - \zz D \theta^1)  +3 e^{-\frac{2}{\sqrt{3}}
\varphi_1}G_1 \wedge *G_1 \nonumber \\
&&
+\tfrac{1}{4}e^{-\sqrt{3}\varphi_1-\varphi_2} F^1_1 \wedge * F^1_1
 +\tfrac{1}{4}e^{-\sqrt{3}\varphi_1+\varphi_2} (F^2_1 - \zz F^1_1)
\wedge * (F^2_1 - \zz F^1_1) \; .
\end{eqnarray}
Recall that $G_1$, $F_1^\alpha$, $\alpha=1,2$, are given in (\ref{G1solvfg}), (\ref{F1fg}), that the covariant derivatives $D \theta^\alpha$, $D\xi^\alpha$ are defined in (\ref{covddoublet}), and that $D \ab$ is defined in (\ref{covDb}). We have also defined
\begin{eqnarray}
&& D \varphi_2 =  d \varphi_2 +2g A_1 \zz \; , \nn
&& D \zz = d \zz +gA_1 (e^{-2\varphi_2}-\zz^2-1) \; .
\end{eqnarray}
In this form, it is now apparent that the Lagrangian (\ref{HMpara}) is equivalent to (\ref{LagHMcan}) with $h_{uv}$ being the metric on $G_{2(2)}/SO(4)$ given in  (\ref{HMg=0metric}), and $q^u$, $u=1, \ldots, 8$, the  scalars given in
(\ref{coordsG2SO4}).

From the definition of the covariant derivatives, we can read off
\begin{eqnarray} \label{KillG2SO4}
k_0 &=& -2\zz \partial_{\varphi_2} -(e^{-2\varphi_2}-\zz^2-1)  \partial_\zz -\theta_2 \partial_{\theta_1} +\theta_1 \partial_{\theta_2} -\tfrac{1}{4} l \partial_\ab -\xi_2 \partial_{\xi_1} +\xi_1 \partial_{\xi_2} \; , \nonumber \\
k_1 &=& \tfrac{1}{2}  \partial_\ab \; ,
\end{eqnarray}
as the Killing vectors $k_I$, $I=0,1$, of $G_{2(2)}/SO(4)$ along which the gauging is turned on. It can be checked that the vectors (\ref{KillG2SO4}) do indeed leave the metric (\ref{HMg=0metric}) invariant.
We have a $U(1) \times \mathbb{R}$ gauging, as noted
at the end of section 4.1. To see this explicitly, we introduce the
$\mathfrak{u}(1) \oplus \mathbb{R} \subset \mathfrak{g}_2$-algebra valued gauge field
\begin{eqnarray}\label{exx}
X_1 = A_1 \textsf{K}_1 + \tfrac{\sqrt{3}}{2} (l A_1 + 2\tilde B_1) \textsf{F}_4
\end{eqnarray}
and use the embedding tensor approach (see \cite{Samtleben:2008pe} for a review)
to write the covariant derivative of the hyperscalars matrix ${\cal M}$ defined in (\ref{hypersmatrix}) as
\begin{equation}
D {\cal M} = d {\cal M} + g( X_1 {\cal M} + {\cal M} X_1^\sharp ) \;
\end{equation}
($\sharp$ being the generalised transpose defined in appendix \ref{hyperMS}) and the Lagrangian (\ref{HMpara}) as
\begin{equation}
{\cal L}_{\textrm{HM}} = \tfrac{1}{16} \Tr \left( {\cal M}^{-1} D {\cal M} \wedge * {\cal M}^{-1}  D {\cal M} \right) \; .
\end{equation}
Note that from \eqref{exx} we deduce that the Killing vector $k_0 +(l/2) k_1$ is associated to the compact generator $\textsf{K}_1 = \textsf{E}_1-\textsf{F}_1$, while $-(1/\sqrt 3)k_1$ is related to the negative root non-compact generator $\textsf{F}_4$.

\subsubsection{Scalar potential}

Given the $Sp(1)$ spin connection $\omega^x$ on $G_{2(2)}/SO(4)$, and its curvature $K^x$, presented in equations (\ref{Sp1spincon}) and (\ref{Sp1curv}), we can work out the momentum maps $P^x_I$ corresponding to the Killing vectors in (\ref{KillG2SO4}) via the definition (\ref{mommapsdef}). For $k_0$ we find
\begin{align}
P_{0}^{1}=& \tfrac{1}{2} e^{\tfrac{1}{2}\vec{\alpha}_{5}\cdot \vec{\varphi}} \left(-\xi_{2}+\theta_{1}\left( \tfrac32 l - \theta^T \theta \right)\right) +\tfrac32 e^{\frac{1}{2}\vec{\alpha}_{3}\cdot \vec{\varphi}} \left(\theta_{1}+\zz\theta_{2} \right) \; , \nonumber\\[8pt]
P_{0}^{2}=& \tfrac{1}{2} e^{\frac{1}{2}\vec{\alpha}_{6}\cdot \vec{\varphi}} \left(\xi_{1}+\zz\xi_{2}+\left(\theta_2 -\zz\theta_{1} \right) \left( \tfrac32 l - \theta^T \theta \right) \right) +\tfrac32 e^{\frac{1}{2}\vec{\alpha}_{2}\cdot \vec{\varphi}}\theta_{2} \; , \nonumber\\[8pt]
P_{0}^{3}=& \tfrac{1}{2} e^{\frac{1}{2}\vec{\alpha}_{1}\cdot \vec{\varphi}}\left(e^{-2 \varphi_2} +\zz^{2}+1\right) -\tfrac34 e^{\frac{1}{2}\vec{\alpha}_{4}\cdot \vec{\varphi}} \left(l-2 \theta^T \theta  \right) \; ,
\end{align}
where $\vec \alpha_i$, $i=1, \ldots, 6$, are the $G_2$ roots given in (\ref{posroots}), $\vec\varphi \equiv (\varphi_1, \varphi_2)$, and a dot denotes Euclidean scalar product. For $k_1$ we have
\begin{eqnarray}
P_{1}^{1}= -\tfrac{3}{2}e^{\frac{1}{2}\vec{\alpha}_{5}\cdot \vec{\varphi}}\theta_{1} \, , \qquad P_{1}^{2}= -\tfrac{3}{2}e^{\frac{1}{2}\vec{\alpha}_{6}\cdot \vec{\varphi}}\,\left(\theta_{2}-\zz \theta_{1}\right) \, , \qquad
P_{1}^{3}= -\tfrac{3}{2}e^{\frac{1}{2}\vec{\alpha}_{4}\cdot \vec{\varphi}} \; .
\end{eqnarray}

Equipped with all these definitions, we can verify, after some calculation, that
the scalar potential of our $D=4$ reduced theory in (\ref{lagPotfgt})
agrees with the canonical $N=2$ expression (\ref{scalarpotcan}):
${\cal L}_\textrm{pot} = -g^2 V \textrm{vol}_4$.

\section{$AdS_4$ vacua and mass spectrum}
In this section we will discuss the $AdS_4$ vacua of the $D=4$ reduced theory described in
the preceding sections.
We find two known $AdS_4$ vacua both of which have $l=-1$ i.e. $\Sigma_3=H^3$ (or $H^3/\Gamma$).
One of these $AdS_4\times H^3$ solutions \cite{Gauntlett:2000ng}
is supersymmetric and after uplifting on $S^4$ to $D=11$
is interpreted as being dual to the SCFT arising on M5-branes wrapping SLag 3-cycle $H^3$. The second
$AdS_4\times H^3$ solution is not supersymmetric and was found in \cite{Gauntlett:2002rv}.
Here we shall recall these solutions and, within the consistent truncation, determine the spectrum of
operators in the dual CFTs.

As mentioned previously, in this and the remaining sections of the paper, we will use
the field variables of section 3, for which the equations of motion are given in
\eqref{firstsetone}-\eqref{4dEinstein}.

\subsection{Supersymmetric $AdS_4$\label{SUSY_vac}}\label{susyvacsection}
The supersymmetric $AdS_4$ solution is obtained by setting $l=-1$,
\bea
e^{-20\phi}=2,\qquad
e^{10\lambda}=2,
\eea
with all other fields trivial, and the $AdS_4$ radius squared $L^2$ is given by
\be\label{susyadssl}
L^2=\frac{{\sqrt 2}}{g^2}\,.
\ee

We now consider the masses of the fields in this vacuum. We find that
the $\phi,\lambda$ fields mix to give masses
\bea
M^2L^2=3\pm{\sqrt{17}}\,,
\eea
corresponding to operators in the dual SCFT with scaling dimensions
\bea
\Delta=\frac{1}{2}+\frac{1}{2}{\sqrt{17}},\qquad
\Delta=\frac{5}{2}+\frac{1}{2}{\sqrt{17}}\,.
\eea
The $\beta$ field doesn't mix and has mass given by
\be
M^2L^2=2
\ee
and hence scaling dimension
\be
\Delta=\frac{3}{2}+\frac{1}{2}{\sqrt{17}}\,.\nn
\ee

We next consider the fields $B_2$, $C_1$ and $A_1$. Using \eqref{firstsettwo}
to solve for $B_2$ and then substituting into \eqref{firstsetone} (at linearised order), and then combining with
\eqref{F2eom} we obtain coupled
equations for two vector fields $C_1$ and $A_1$.
After diagonalisation these give masses
\be
M^2L^2=0,4
\ee
and hence
\bea
\Delta=2,\qquad \Delta=\frac{3}{2}+\frac{1}{2}{\sqrt{17}}\,.
\eea
Note that the massive mode is given by $C_1$ while the massless mode is given by the combination $A_1+(3/g)C_1$ and corresponds to
the abelian $R$-symmetry current of the dual $N=2$ SCFT.

We now turn to the charged scalar fields.
We first consider the fields $\chi$, $\theta$ and $h_3$. After using
\eqref{firstsetfour} to solve for $h_3$ and then substituting into \eqref{firstsetthree} we obtain
coupled equations for $\chi$, $\theta$. After taking suitable linear combinations
of $\chi$ and $\epsilon\theta$ we can diagonalise the mass matrix leading to masses
\bea
M^2L^2=10, 2
\eea
and hence scaling dimensions
\bea
\Delta=5, \qquad \Delta= \frac{3}{2}+\frac{1}{2}{\sqrt{17}}\,.
\eea
From \eqref{covddoublet} we observe that these scalars have $SO(2)$ charge $g$.
We next analyse the two scalar degrees of freedom in $\cT$. To do so it will be useful
to now choose the explicit parametrisation of $SL(2,\mathbb{R})/SO(2)$ given by\footnote{Changing between the parametrisations (\ref{Npar}) and (\ref{nparam}) of $\cT$ is equivalent to the change of coordinates $e^{\varphi_2} = \cosh \rho + \cos \sigma \sinh \rho$ and $e^{\varphi_2} \zeta = \sin \sigma \sinh \rho$ on $SL(2,\mathbb{R})/SO(2)$.}
\begin{align}\label{nparam}
\cT \left[\rho,\sigma \right]=&R\left[\sigma\right]^{-1}
\left(   \begin{matrix}
      e^{\rho} & 0 \\
      0 & e^{-\rho} \\
   \end{matrix}\right)
R\left[\sigma\right]\,,\nonumber\\
R\left[\sigma\right]=&\left(
   \begin{matrix} 
      \cos\left(\frac{\sigma}{2}\right) & \sin\left(\frac{\sigma}{2}\right) \\
      -\sin\left(\frac{\sigma}{2}\right) & \cos\left(\frac{\sigma}{2}\right) \\
   \end{matrix}\right)\,,
\end{align}
where $\sigma$ is a periodic coordinate with period $2\pi$ and $\rho>0$ since
$\cT \left[-\rho,\sigma \right]= \cT \left[\rho,\sigma+\pi \right]$. Using this
we find that the corresponding kinetic term in the Lagrangian \eqref{4dLagKin} can be written as
\begin{equation}
\frac{1}{4}\Tr(\cT^{-1}D\cT \wedge * \cT^{-1}D\cT) =\frac{1}{2}\left[d\rho\wedge\ast d\rho +\sinh^{2}\rho \left(d\sigma-2gA_1 \right) \wedge \ast\left(d\sigma-2gA_1 \right)\right]\,.
\end{equation}
After expanding about the supersymmetric $AdS_4$ vacuum we find a complex scalar field
with mass given by
\be
M^2L^2=4
\ee
and hence scaling dimension
\be\label{dimnmode}
\Delta=4\,.
\ee
Using \eqref{covddoublet} we also observe that the $R$-charge of the complex scalar in $\cT$ is $2g$ i.e.
twice that of the complex scalar degree of freedom in $\theta$ and $\chi$.
Note that the above scalar operators in the dual SCFT,
except one coming from the $\phi,\lambda$ sector, and the massive vector are all irrelevant ($\Delta>3$).

By considering the conformal dimensions of the fields and their $R$-charges we can now arrange these
into $OSp(2|4)$ multiplets (see e.g. \cite{fab}). The graviton ($\Delta=2$) and the massless vector
($\Delta=3$) are both neutral and form a massless graviton multiplet (see table 8 of \cite{fab}).
The complex scalar in $\chi,\theta$ with $\Delta=5$ and R-charge one (in units of $g$) combined with
the complex scalar in $\cT$ with $\Delta=4$  and R-charge two form a hypermultiplet
(see table 7 of \cite{fab}). The remaining fields, three neutral scalars with $\Delta=E_0, E_0+1, E_0+2$,
one complex scalar with $\Delta=E_0+1$ and unit charge and the massive neutral vector with
$\Delta=E_0+1$, where $E_0=(1+{\sqrt{17}})/2$ form a long vector multiplet (table 3 of \cite{fab}).
Note that since the spectrum contains irrational scaling dimensions the abelian $R$-symmetry group
of the SCFT is a non-compact $\mathbb{R}$.

\subsection{Non-susy vacuum}
We now consider the non-supersymmetric $AdS_4$ solution first found in \cite{Gauntlett:2002rv}.
This is obtained by setting $l=-1$,
\bea
e^{-20\phi}=\frac{486}{625},\qquad
e^{10\lambda}=10
\eea
and the $AdS_4$ radius squared $L^2$ is given by
\be
L^2=\frac{5{\sqrt 2}}{3{\sqrt 3}}\frac{1}{g^2}\,.
\ee

We next discuss the mass spectrum about this vacuum.
The $\phi,\lambda$ fields mix and give masses
\bea
M^2L^2=\frac{23}{5}\pm\frac{1}{5}{\sqrt {409}}
\eea
corresponding to scaling dimensions
\bea
\Delta&=&\frac{3}{2}+\frac{1}{10}[685\pm20{\sqrt{409}}]^{1/2}\,.
\nn
\eea
The $\beta$ field has mass given by
\be
M^2L^2=\frac{6}{5}
\ee
corresponding to
\be
\Delta=\frac{3}{2}+\frac{1}{10}{\sqrt{345}}\,.
\ee

For the $B_2$, $C_1$ and $A_1$ fields, by again solving for $B_2$ and then considering the linearised
equations for $A_1$ and $C_1$ we are led to two vectors with
masses
\be
M^2L^2=0,28/5
\ee
and hence
\bea
\Delta=2,\qquad\Delta=\frac{3}{2}+\frac{3}{10}{\sqrt{65}}\,.
\eea
The combination $A_1+(27/7g)C_1$ is the massless mode and $C_1$ is the massive mode.

Finally we consider the charged fields.
After eliminating $h_3$ we again find that $\epsilon\theta,\chi$ mix to give a complex field with
mass
\bea
M^2L^2=\frac{134}{5}\pm\frac{4}{5}{\sqrt{241}}
\eea
and hence
\bea
\Delta=\frac{3}{2}+\frac{1}{10}{\sqrt {2905\pm 80{\sqrt{241}}}}\,.
\eea
Using the parametrisation of $\cT$ given in \eqref{nparam}
at linearised order we find a complex scalar field with mass
\be
M^2L^2=68
\ee
and hence
\be
\Delta=\frac{3}{2}+\frac{1}{2}{\sqrt{281}}\,.
\ee
Note that all of the above scalar fields and also the massive vector are dual to irrelevant operators in the dual CFT.

\subsection{Additional $AdS_4$ vacua?}
In searching for additional $AdS_4$ vacua, we must impose that
$C_1=B_2=A_1=h_3=0$ and that all scalar fields are constant. Then \eqref{firstsetthree} immediately implies that $\chi=0$.

We now show there are no additional $AdS_4$ solutions when $l=0$ or $l=-1$.
From \eqref{betaeom} we deduce that $\beta=0$.
Next, \eqref{thetaeom} can be written
\be
\left[e^{-4\phi+8\lambda}(-l+2\theta^T\theta)+\cT^{-1}(e^{5\lambda}\cT-e^{-5\lambda})^2\right]\theta=0
\ee
which implies (for $l=0,-1$) that $\theta=0$.
We then just have the $\phi,\lambda,\cT$ system. Next using the parametrisation of $\cT$ given in \eqref{nparam},
we deduce from \eqref{enneom} that, without loss of generality, we can take $\cT$ to be diagonal.
From \eqref{enneom} we immediately deduce that either $\rho=0$ or $\cosh\rho=(3/2)e^{-10\lambda}$.
In the former case we easily conclude that there is just the supersymmetric and non-supersymmetric $AdS_4$ solutions
discussed above \cite{Gauntlett:2002rv}. In the latter case we find that
\eqref{lambda_eom2} and \eqref{4dbreathing} imply that
\bea
x^2l^2+16-8y&=&0\,,\nn
40xl-7x^2l^2+48+16y&=&0,
\eea
respectively, where we have defined $x\equiv e^{8\l-4\phi}, y\equiv e^{20\l}$. It is now
simple to see that there are no solutions when $l=0$. When $l=-1$ these equations have a
positive solution, but it gives rise to a complex value of $\rho$ and hence there
are no additional $AdS_4$ solutions when $l=-1$ either (in the $\phi,\lambda,\rho$ sector this was already stated in \cite{Gauntlett:2002rv}).

The above analysis also implies that there are no additional solutions when $l=+1$ and we impose
$\beta=\theta=0$.  If we take $\theta=0$ and $\beta\ne 0$ we see from \eqref{betaeom} that $\beta^2=1/2$ and following
similar arguments we again find no additional solutions. This just leaves open the possibility of $AdS_4$ solutions with $l=+1$ and $\theta\ne 0$,
which we will not address here.

\section{Additional consistent truncations}\label{addcontrunc}
In this section we shall discuss some additional truncations of the consistent KK truncation that
we presented in section \ref{conkktrunc}. We make no attempt to be comprehensive.

\subsection{Minimal gauged supergravity - Einstein-Maxwell theory\label{EM}}
\label{emkkred}
The supersymmetric $AdS_4$ vacuum discussed in section \ref{susyvacsection}
is a specific example of the general class of
supersymmeric $AdS_4$ solutions of $D=11$ supergravity, dual to $N=2$ SCFTs in $d=3$,
that were classified using $G$-structure techniques in \cite{Gauntlett:2006ux}.
For any solution in this general class, it has already been shown that there is a consistent truncation
to minimal gauged supergravity, with bosonic fields consisting of a metric and a gauge-field
\cite{Gauntlett:2007ma}.
Thus we should be able to further truncate the ansatz
in \eqref{KKmetricf}-\eqref{KKscalarsf} to obtain the bosonic content of minimal gauged supergravity.
This is simple to do.

We set
\begin{align}
l=-1,\qquad
e^{-20\phi}=2,\qquad
e^{10\lambda}=2\,,
\end{align}
as in the supersymmetric $AdS_4$ vacuum
and also
\be
B_2=-\frac{1}{\sqrt 2}*F_2\,.
\ee
Finally, we set
$C_{1}$, $\chi$, $\theta$, $\cT$, $h_{3}$ and $\beta$ to their trivial values.
We then find that all equations of motion \eqref{firstsetone}-\eqref{4dEinstein} boil down to
\bea
R_{\mu\nu}&=&-\frac{3g^2}{\sqrt{2}}\,g_{\mu\nu}+\frac{1}{\sqrt{2}}\,\left(F_{\mu\rho}F_{\nu}{}^{\rho}-\frac{1}{4}g_{\mu\nu}F_{\rho\sigma}F^{\rho\sigma} \right)\nn
d*F&=&0\,.
\eea
These equations of motion come from the bosonic Lagrangian of minimal gauged supergravity,
which is just Einstein-Maxwell theory with a negative cosmological constant:
\begin{eqnarray} \label{4dLagKinr}
{\cal L}_\textrm{kin} &=& [R^{(4)} +\frac{6}{L^2}]\textrm{vol}_4 -\tfrac{1}{\sqrt{2}}F_2 \wedge * F_2\,,
\end{eqnarray}
where (c.f. \eqref{susyadssl})
\be
L^2=\frac{\sqrt 2}{g^2}\,.
\ee

For example, one solution is the standard electrically charged
AdS Reissner-N\"ordstrom black hole with flat spatial sections (also called a black brane) given by
\begin{align}\label{adsrn1}
ds^{2}=&-f\,dt^{2}+\frac{dr^{2}}{f}+\frac{r^2}{L^2}\,\left(dx_{1}^{2}+dx_{2}^{2} \right)\,,\nn
F_{2}=&\frac{\mu_{e} r_{+}}{r^{2}}\,dt\wedge dr\,,
\end{align}
with
\begin{equation}\label{adsrn2}
f=\frac{r^2}{L^2}-\left(\frac{r_{+}^{2}}{L^2}+\frac{\mu_{e}^{2}}{2\sqrt{2}} \right)\,\frac{r_{+}}{r}+\frac{\mu_{e}^{2}}{2\sqrt{2}}\frac{r_{+}^{2}}{r^{2}}\,.
\end{equation}
This solution, after uplifting to $D=11$,
describes the SCFT on M5-branes wrapped on Slag 3-cycles, $H^3/\Gamma$, when held at finite
temperature and finite chemical potential. The stability of these black holes will be discussed in
section \ref{holsc}.

\subsection{Charged fields to zero}\label{chftz}
It is also consistent with the full equations of motion \eqref{firstsetone}-\eqref{4dEinstein} to set
\be
\chi=h_3=\theta =0\,.
\ee
It is also consistent to then, in addition, set
\be
\cT=\delta\,.
\ee
This latter truncation sets all of the fields carrying non-zero $SO(2)$ charge to zero.

\section{M5-branes wrapping SLag 3-cycles at finite $T,\mu$ }
\label{holsc}
We now use the results obtained so far to initiate a study of the $N=2$ $d=3$ SCFTs, dual to the
supersymmetric $AdS_4\times H^3/\Gamma\times S^4$ solutions, when held at finite temperature $T$ and
chemical potential $\mu$ with respect to the global $R$-symmetry.
As we have already mentioned these SCFTs arise on M5-branes wrapped on SLag 3-cycles
$H^3/\Gamma$ in Calabi-Yau three-folds. The conformal invariance implies that the system
will just depend on the dimensionless parameter $T/\mu$.

At high temperatures the system is described by the (uplifted) electrically charged AdS-RN black hole
that was given in \eqref{adsrn1}-\eqref{adsrn2}. In this section we will investigate the possibility that
the AdS-RN black hole has unstable linearised modes below given ``branching temperatures". At a branching temperature the
corresponding linearised mode becomes a zero mode and indicates that a new branch of black hole solutions
is appearing. We will see that there are two new types of black hole branches, one with charged hair, corresponding
to holographic superconductivity, and the other without. We will see that the branching temperature
of the superconducting black holes is lower than that of the other branch. In order to determine
which is thermodynamically preferred one will need to go beyond the linearised analysis that
we perform here to determine the order of the phase transition. Specifically, if the transition to the superconducting
black holes is first order, the ``critical temperature" at which the system moves, discontinuously, from the AdS-RN black holes to the
superconducting branch is higher than that of the superconducting black hole branching temperature\footnote{Some
general discussion of related issues appear in \cite{Kol:2006vu} and some bottom up examples of holographic superconductors with first order transitions are described in \cite{Franco:2009yz}\cite{Franco:2009if}.} and could be higher than
the critical temperature for the neutral black holes. Furthermore, there could
be additional black hole branches either within or outside the $D=4$ truncation, which could be associated with
even higher critical temperatures.

The simplest way to look for new branches of $D=4$ black hole solutions is to study the zero temperature,
near horizon $AdS_2\times \mathbb{R}^2$ limit of the AdS-RN black hole and look for modes that violate the
$AdS_2$ BF bound  \cite{Hartnoll:2008kx}\cite{Gubser:2008pf}\cite{Denef:2009tp}.
Using this approach in section \ref{simanal}, we find that there are indeed charged modes that violate
the $AdS_2$ BF bound indicating holographic superconductivity. However, we also find some neutral modes
that violate the $AdS_2$ BF bound. This indicates that there are two new branches of black hole solutions emerging.
By a more careful analysis in section \ref{morecomanal}, we will show that non-superconducting black holes
have a higher branching temperature.

\subsection{Instabilities of the $AdS-RN$ black hole at $T=0$}\label{simanal}
The near horizon limit of the $T=0$ AdS RN black hole \eqref{adsrn1}-\eqref{adsrn2}
gives the $AdS_2\times \mathbb{R}^2$ solution
\bea
ds_{4}^{2}&=&{L_{(2)}^{2}}\,ds^{2}\left(AdS_{2} \right)+dx_{1}^{2}+dx_{2}^{2}\,,\nn
F_{2}&=&q\, \mathrm{Vol}\left(AdS_{2}\right)\,,\nn
B_{2}&=&\frac{q}{\sqrt{2}L_{(2)}^2}\,dx_{1}\wedge dx_{2}\,,
\eea
where
\begin{align}
L_{(2)}^2=\frac{1}{3\sqrt{2}g^2},\qquad q=&\frac{1}{\sqrt{3}g}\,.
\end{align}
We now consider some linearised fluctuations about this solution.
We will consider various perturbations that are independent of the coordinates on
$\mathbb{R}^2$ and look for perturbations whose $AdS_2$ mass violates the BF bound.
For a unit radius
$AdS_2$ space this condition is
\be
M^2<-\frac{1}{4}\,.
\ee
We will not consider any perturbations of the metric, but
we have checked that, at linearised order, the perturbations that we
consider do not source any metric perturbations: specifically we have checked that the right hand side of
\eqref{4dEinstein} vanishes at leading order.

We first consider fluctuations in the $h_3,\chi,\theta$ sector. After eliminating $h_3$ we find
that $\chi$ and $\epsilon\theta$ mix, exactly as in the $AdS_4$ vacuua studied in
section \ref{susyvacsection}. At linearised order we find
\bea
-D*_2D\chi+\frac{3}{2}D*_2D(\epsilon\theta)-2\sqrt{2}g^2\chi {\rm Vol}(AdS_2)&=&0\,,\nn
D*_2D(\epsilon\theta)+g^2[\sqrt{2}\chi+\frac{5}{\sqrt 2}(\epsilon\theta)]{\rm Vol}(AdS_2)&=&0\,.
\eea
This gives $AdS_2$ masses equal to $5/3$ and $1/3$.
Hence there is no instability in this sector.

We next consider the scalars parametrising the $SL(2,\mathbb{R})/SO(2)$ coset, $\cT$.
Using the parametrisation \eqref{nparam} we consider
small fluctuations of $\rho$.
After checking that
it is consistent with the equations of motion, we set $\sigma=0$ and
then find that at linearised order $\rho$ decouples and satisfies
\begin{equation}
\Box_{2}{\rho}+\frac{2}{3}{\rho}=0\,,
\end{equation}
where $\Box_2$ is the Laplacian on a unit radius $AdS_2$.
This gives $M^2=-2/3$ which violates the $AdS_2$ BF bound.
This shows that the system is unstable to condensing a charged mode.
As we will discuss in the next subsection we can use the results of \cite{Denef:2009tp} to argue that
there will be a new branch of black holes that will spontaneously break the abelian R-symmetry and
hence exhibit holographic superconductivity.

We now consider fluctuations of the neutral scalars that mix with some fluctuations of the gauge fields.
Specifically, for the scalars we consider
\bea
\beta&=&\delta\beta\,,\nn
\lambda&=&\frac{1}{10}\ln 2+\delta{\lambda}\,,\nn
\phi&=&-\frac{1}{20}\ln 2+\delta{\phi}
\eea
and for the gauge-fields
\begin{align}
F_{2}=&\left(q+\delta{F}\right)\, \mathrm{Vol}\left(AdS_{2}\right)\,,\nn
B_{2}=&\left(\frac{qR^{2}}{\sqrt{2}}+\delta{B} \right)\,dx_{1}\wedge dx_{2}+\delta{B}'\, \mathrm{Vol}\left(AdS_{2}\right)\,,
\end{align}
with, from \eqref{firstsetone},
\be
C_1=-\frac{1}{2g}*_2d(\delta B)\,.
\ee
In order to solve the equations of motion \eqref{firstsetfour}, \eqref{F2eom} at leading order, we quickly see that these
fluctuations are not independent. We find that we should set
\bea
\delta B'&=&2q\delta\beta\,,\nn
\delta F&=&-3\sqrt{2}R^2_{(2)}\delta B-6q(\delta\phi-2\delta\lambda)\,.
\eea
Substituting into the rest of the equations we find that the mode $\delta\beta$ decouples from
the others and satisfies
\be
\Box_2\delta\beta-\frac{4}{3}\delta\beta=0
\ee
and thus does not give rise to any instability.
The remaining modes remain coupled and satisfy
\begin{align}\label{ceqns1}
\Box_{2}(q\delta{B})-\frac{2}{3}(q\delta{B})-\frac{4}{3}\left(\delta{\phi}-2\delta{\lambda} \right)=&0\,,\nn
\Box_{2}\delta{\lambda}-\frac{2}{3}\delta{\lambda}+\frac{2}{3}\delta{\phi}+\frac{2}{5}(q\delta{B})=&0\,,\nn
\Box_{2}\delta{\phi}+\frac{2}{3}\delta{\lambda}-\frac{4}{3}\delta{\phi}-\frac{1}{5}(q\delta{B})=&0\,,
\end{align}
yielding the mass spectrum on $AdS_{2}$
\begin{equation}
M^2=\frac{1}{3}\left(3-\sqrt{17}\right) ;\quad \frac{1}{3}\left(3+\sqrt{17} \right);\quad\frac{2}{3}\,.
\end{equation}
Thus the $F_2,B_2,C_1,\phi,\lambda$ sector also contains an unstable mode.
Recall from section 5.1 that in the supersymmetric $AdS_4$ vacuum the two neutral scalars $\phi,\lambda$
are dual to one relevant operator (with $\Delta\approx 2.56$)
and one irrelevant operator, and the $B_2,C_1$ fields describe a neutral
massive vector which is dual to an irrelevant operator. The detailed interactions between these
fields and also with $F_2$ give rise to the violation of the $AdS_2$ BF bound (c.f. the simpler mechanism
involving a single neutral scalar field in a bottom up setting discussed in \cite{Hartnoll:2008kx}).

We have thus found two unstable modes. One is charged and comes from the $\cT$ sector while the
other is neutral and
comes from the $F_2,B_2,C_1,\phi,\lambda$ sector. Each of these unstable modes
will give rise to a new branch of charged black holes
that will appear at some branching temperature, the former with charged hair and the latter with neutral hair.
To determine the branching temperatures we need
to consider the linearised fluctuations about the electrically charged $AdS$-RN black hole at non-zero temperature.

\subsection{Instabilities of the AdS-RN black hole at $T\ne 0$}\label{morecomanal}

We now study the perturbative stability of the finite temperature electrically charged $AdS$
RN black hole of section \ref{EM}, focussing on the modes associated with those violating the
$AdS_2$ BF bound in the zero temperature limit that we identified in the last subsection.
We begin by writing the $AdS$ RN black hole metric and vector potential \eqref{adsrn1}-\eqref{adsrn2}
as
\bea\label{dord}
ds_{4}^{2}&=&-f\,dt^{2}+\frac{dr^{2}}{f}+\frac{r^{2}}{L^{2}}\,\left(dx_{1}^{2}+dx_{2}^{2}\right)\,,\nn
F_2&=&\frac{F_0}{r^2}Vol_2\,,\nn
B_2&=&\frac{F_0}{\sqrt{2}L^2}Vol_{\mathbb{R}^2}\,,
\eea
where ${\rm Vol}_2=dt\wedge dr$, ${\rm Vol}_{\mathbb{R}^2}=dx^1\wedge dx^2$ and
\begin{align}
f=&\frac{r^{2}}{L^2}-\left(\frac{r_{+}^{2}}{L^2}+\frac{\mu^{2}}{2\sqrt{2}} \right)\,\frac{r_{+}}{r}+\frac{\mu^{2}}{2\sqrt{2}}\frac{r_{+}^{2}}{r^{2}}\,,\nn
F_0=&-\mu r_+,\qquad L^{2}=\frac{\sqrt{2}}{g^{2}}\,.
\end{align}

We will consider modes which are functions of $t,r$ only. For the $\cT$ sector, in the parametrisation
\eqref{nparam} we again consider
a perturbation $\delta\rho$ with $\sigma=0$. Using that $F_2=dA_1$ with $A_1=Adt$,
$A=\mu\,\left(1-\frac{r_{+}}{r}\right)$ we find that $\delta\rho$ must satisfy
\begin{align}
&\Box_{4}\delta{\rho}+\frac{4 g^2A^{2}}{f}\delta{\rho}-2\sqrt{2}g^{2}\delta{\rho}=0\label{rhoadsrn}\,,
\end{align}
where $\Box_4$ is the $D=4$ Laplacian for the metric in \eqref{dord}.

We next consider the $F_2,B_2,\phi,\lambda$ sector.
For the scalars we consider
\bea
\lambda&=&\frac{1}{10}\ln 2+\delta{\lambda}\,,\nn
\phi&=&-\frac{1}{20}\ln 2+\delta{\phi}\,.
\eea
For the two forms we will take
\begin{align}
F_{2}=&\left[\frac{F_{0}}{r^{2}}+\delta{F}\right]\mathrm{Vol}_{2}\,,\nn
B_{2}=&\left[\frac{F_{0}}{\sqrt{2}L^2}+\delta{B}\right]\mathrm{Vol}_{\mathbb{R}^{2}}\,,
\end{align}
with
\bea
C_{1}&=&-\frac{1}{2g}\ast_{4}d(\delta B)\,,\nn
\delta{F}&=&-\frac{3\sqrt{2}L^{2}}{r^{2}}\delta{B}-\frac{6F_{0}}{r^{2}}\left(\delta{\phi} -2\delta{\lambda}\right)\,.
\eea

We then find that the equation \eqref{firstsettwo} reads
\begin{align}
L^{2}r^{2}d\left(\frac{1}{r^{2}}\ast_{2}d(\delta{B})\right)+\frac{4}{\sqrt{2}}g^{2}L^{2}\delta{B}+4F_{0}g^{2}\left(\delta{\phi}-2\delta{\lambda}\right)=&0\,.\label{B2eom}
\end{align}
After defining
\begin{equation}
\delta{B}=r^{2}\frac{F_{0}}{L^{2}r_{+}^{2}}\delta{b}\,,
\end{equation}
we find that this equation and all remaining equations reduce to the coupled system
\begin{align}
-\Box_{4}\delta{b}+2\left(-\partial_{r}\left(\frac{f}{r}\right)+\sqrt{2}g^{2} \right)\delta{b}+\frac{4g^{2}r_{+}^{2}}{r^{2}}\left(\delta{\phi}-2\delta{\lambda} \right)=0\,,\nn
-\Box_{4}\delta{\lambda}-g^{2}\frac{2\sqrt{2}}{5}\left(\delta{\lambda}+2\delta{\phi} \right)-\frac{2F^{2}_{0}}{5}\left(\frac{1}{r^{2}r_{+}^{2}}\delta{b}+\frac{1}{r^{4}\sqrt{2}}\left(\delta{\phi}-2\delta{\lambda} \right)\right)=&0\,,\nn
-\Box_{4}\delta{\phi}+g^{2}\frac{\sqrt{2}}{5}\left(-4\delta{\lambda}+17\delta{\phi} \right)+\frac{F^{2}_{0}}{5}\left(\frac{1}{r^{2}r_{+}^{2}}\delta{b}+\frac{1}{r^{4}\sqrt{2}}\left(\delta{\phi}-2\delta{\lambda} \right)\right)=&0\,.\label{tphieom}
\end{align}
One can check that upon setting $r=r_+$, one recovers the $AdS_2$ equations
given in \eqref{ceqns1} (after rescaling $\delta b$).

By numerically solving \eqref{rhoadsrn} and \eqref{tphieom} we can determine
the temperatures at which the new branches of black hole solutions appear.
Since we are looking for zero modes we consider perturbations that are independent
of time (i.e. $e^{-i\omega t}$ with $\omega=0$). The modes we are interested in are independent
of the spatial coordinates and so the modes just depend on $r$. We then expand out near the horizon and
integrate out to infinity, looking for the temperature at which the non-normalisable asymptotic behaviour vanishes.
In fact for \eqref{rhoadsrn} this analysis has already been performed by Denef and Hartnoll in \cite{Denef:2009tp}.
In their notation we have\footnote{In more detail, we
should set $M_{DH}^2=2$, $g^2_{DH}=1/\sqrt{2}$, $L^2_{DH}=\sqrt{2}/g^2$, $\gamma^2_{DH}=4/g^2$, $q_{DH}=2g$.}
$\gamma_{DH}q_{DH}=4$ and, as we showed in \eqref{dimnmode}, $\Delta=4$. We have solved the numerical problem and we find
the critical temperature $\gamma_{DH}T_c/\mu\approx .001$, which agrees with Figure 1 of \cite{Denef:2009tp}.
We also numerically solved \eqref{tphieom} and find $\gamma_{DH}T_c/\mu\approx .0045$.

Thus, in conclusion, we have shown that as we lower the temperature of the AdS RN black hole, two new branches of black holes appear. The first branch that appears are a new class of charged black holes
carrying neutral scalar and massive vector hair. At a lower temperature a second branch of charged black holes
appear carrying charged scalar hair which spontaneously break the $R$-symmetry and hence
exhibit holographic superconductivity. It would be interesting to determine if the phase
transitions are first or second order. It would also be interesting to construct the fully back reacted
thermodynamically preferred black hole solutions and study their behaviour at lower temperatures.

\section{Lifshitz Solutions}
In this section we investigate the possibility that the equations of motion of the
$D=4$ reduced theory, given in \eqref{firstsetone}-\eqref{4dEinstein}), admits Lif$_4$($z$) solutions.
After uplifting to $D=11$ such solutions would be dual to $d=3$ field
theories with Lifshitz scaling and dynamical exponent $z$. After reducing the problem to solving
a set of algebraic equations we find (using Mathematica) one solution with $z=39.05...$.

For simplicity we restrict our analysis to the truncation where
$\chi=h_3=\theta=0$, discussed in section \ref{chftz}, and consider the following ansatz.
For the metric we take
\begin{align} \label{Lifmetric}
ds^{2}_{4}=&-\frac{r^{2z}}{L^{2z}}\,dt^{2}+\frac{L^{2}}{r^{2}}\,dr^{2}+\frac{r^{2}}{L^{2}}\,\left(dx_{1}^{2}+dx_{2}^{2}\right)\,,
\end{align}
which is the standard Lif$_4$($z$) metric in $D=4$.
We also take
\begin{align}
A_{1}=& q\,\frac{r^{z}}{L^{z}}dt\,,\nn
C_{1}=& -{c}\,\frac{r^{z}}{L^{z+1}}dt\,,\nn
B_{2}=&b\frac{r^2}{L^3}dx_{1}\wedge dx_{2}\,,
\end{align}
where $q,c,b$ are constants, and all remaining scalar fields are taken to be constant.
Note that the scaling symmetry \eqref{scalingsymmetry} can be used to set $L=1$ if desired.
We observe that this ansatz is consistent with the Lif$_4(z)$ scaling symmetry
\begin{equation}
 t\rightarrow s^{z}t,\quad x_{i}\rightarrow s x_{i},\quad r\rightarrow s^{-1}r\,,
\end{equation}
where $z$ is the (constant) dynamical exponent.
One can check that it is consistent with the $D=4$ equations of motion
to now further set $\beta=0$ and, in the parametrisation of $\cT$ given in \eqref{nparam},
$\sigma=0$, and we shall do so for additional simplicity.
The ansatz is thus specified by eight constants: $q,b,c,z,\rho,\lambda,\phi$ and $gL$.

Substituting this ansatz into \eqref{firstsetone} and \eqref{firstsettwo} we get
\begin{align}
&c=\frac{e^{4\lambda+8\phi}}{gL}b\label{c_equation}\,,\\
&2\left[(gL)^2-2ze^{8\lambda+6\phi}\right]b=(gL)^2lqze^{4\lambda-2\phi}\,.\label{b_equation}
\end{align}
From the equation of motion for  the gauge field \eqref{F2eom} we obtain
\begin{equation}
2qze^{-12\lambda+6\phi}-4(gL)^{2}q\sinh^{2}\rho+3lb=0\,.\label{lif_bianchi}
\end{equation}
From the $\lambda$, $\phi$ and $\rho$ equations of motion, \eqref{lambda_eom2}-\eqref{4dbreathing},
and using \eqref{c_equation},  we obtain
\bea
&(gL)^2\left[l^2 e^{20\lambda}+4e^{4\lambda+8\phi}\left(1+2e^{20\lambda}\sinh^2\rho-e^{10\lambda}\cosh\rho\right)\right]\nn
&+2q^2z^2e^{20\phi}-e^{8\lambda+16\phi}\left[2-\frac{8}{(gL)^2} e^{8\lambda+6\phi}\right]b^2=0\,,\label{lambdaeq}\\
&
(gL)^2\left[10 l e^{12\lambda+4\phi}-\frac{7}{4}l^2 e^{20\lambda}+e^{4\lambda+8\phi}
\left(3-4e^{20\lambda}\sinh^2\rho+12e^{10\lambda}\cosh\rho\right)\right]\nn
&-q^2z^2e^{20\phi}+e^{8\lambda+16\phi}\left[1+\frac{16}{(gL)^2} e^{8\lambda+6\phi}\right]b^2=0\,,\label{phieq}\\
&\sinh\rho\,\left[-3+2e^{10\lambda}\cosh\rho -2q^2e^{-2\lambda+6\phi}\cosh\rho \right]=0\,.\label{rhoeq}
\eea

We now turn to Einstein's equations \eqref{4dEinstein}. Observe that for
our metric ansatz the non-zero components for the Ricci tensor are given by
\begin{align}
R_{tt}=\frac{z\left(z+2\right)r^{2z}}{L^{2+2z}},\qquad
R_{rr}=-\frac{z^{2}+2}{r^{2}},\qquad
R_{ij}=-\frac{\left(z+2\right)r^2}{L^{4}}\,\delta_{ij}\,,
\end{align}
where $i,j=1,2$. We then find that
the $(tt)$, $(rr)$ and $(ii)$ components of \eqref{4dEinstein}  give:
\bea
{z\left(z+2\right)}&=&A+B-C\,,\nonumber\\
-{\left(z^{2}+2\right)}&=&-B+C\,,\nonumber\\
-{\left(z+2\right)}&=&B+C\,,\label{lif_einstein}
\eea
where, again using \eqref{c_equation},
\begin{align}
A=&2(gL)^{2}q^{2}\sinh^{2}\rho +\frac{6}{(gL)^2}e^{4\lambda+8\phi}b^2\,,\nn
B=&\frac{q^2z^2}{4}e^{-12\lambda+6\phi}+\frac{3}{4}e^{-4\lambda+2\phi}b^{2}\,,\nn
C=&-\frac{(gL)^{2}}{2}\left[3l e^{-10\phi}-\frac{3}{8}e^{8\lambda-14\phi}l^2+\frac{1}{2}e^{-6\phi-8\lambda}\left(3-4e^{20\lambda}\sinh^{2}\rho+12e^{10\lambda}\cosh\rho \right) \right]\,.
\end{align}
The three equations \eqref{lif_einstein} can be rewritten as
\bea
A+2B+2C+6&=&0\label{C_solution}\,,\nn
A^2+2A-8B&=&0\label{L2_solution}
\eea
and
\bea
z&=&\frac{4B}{A}\,.\label{z_solution}
\eea
We now observe that \eqref{z_solution} is actually already implied by the previous equations \eqref{c_equation}, \eqref{b_equation} and \eqref{lif_bianchi}.

To summarise, $c$ can be obtained from \eqref{c_equation}. If we assume that
$(gL)^2\ne 2ze^{8\lambda+6\phi}$, as we shall do, then
$b$ can be obtained from \eqref{b_equation}.
This leaves six algebraic equations to be solved, \eqref{lif_bianchi}-\eqref{rhoeq} and
\eqref{C_solution},\eqref{L2_solution}, for six remaining constants
$\phi$, $\lambda$, $\rho$, $q$, $z$ and $(gL)$.
Using Mathematica we found one solution with $l=-1$ and
\begin{align}
z=& 39.059617\dots\nn
gL=& 19.592485\ldots\nn
\lambda=&0.068678\ldots\nn
\phi=&0.043883\ldots\nn
\rho=&0.299400\ldots\nn
q=&-0.907857\ldots
\end{align}
This solution uplifts to a solution of $D=11$ supergravity that is a product of
a Lif$_4(z\sim 39)$ factor with an $H^3\times S^4$ factor, with the latter fibred over the former (due to the
fact that $A_1=(qr^z/L^z) dt$).
It would be interesting to explore this solution
further. It would also be interesting to know if the $D=4$ equations of motion admit further
Lif$_4(z)$ solutions.

\section{Discussion}
We have presented a new consistent KK reduction of $D=11$ supergravity on
$\Sigma_3\times S^4$, where $\Sigma_3=H^3/\Gamma,S^3/\Gamma,R^3/\Gamma$,
to obtain $N=2$ gauged supergravities in $D=4$. For the case of
$H^3/\Gamma$, the $D=4$ theory admits a supersymmetric $AdS_4$ vacuum
which uplifts to a $D=11$ solution dual to the $d=3$ $N=2$ SCFT arising on
M5-branes wrapping SLag 3-cycles $H^3/\Gamma$. We showed that
the $D=4$ theory also admits another non-supersymmetric $AdS_4$ solution
as well as a Lif$_4(z)$ solution with $z\sim 39$. It would be interesting to determine whether
or not there are additional Lifshitz solutions.
It would also be interesting to
investigate whether or not there are $D=4$ domain wall type solutions interpolating
between these solutions that would describe dual RG flows between the different critical points.

We also studied the $N=2$ SCFT arising on the wrapped M5-branes at finite temperature
and chemical potential by studying black holes. The high temperature limit is
described by an uplifted $D=4$ AdS-RN type black hole. We also showed that these black
holes have two instabilities corresponding to the existence of two new branches
of black hole solutions. One branch, the new charged black holes with neutral hair,
preserve the abelian $R$-symmetry and arise from an instability involving two neutral scalars
and a massive vector field.
The other branch of black holes spontaneously break the $R$-symmetry and thus comprise
a new class of holographic superconducting
black holes. We showed that the branching temperature of the charged black holes with neutral hair
is higher than that of the superconducting black holes. Therefore the charged black holes with neutral hair
will be thermodynamically preferred
unless the superconducting black hole transition is first order with a
critical temperature sufficiently higher than its branching temperature. We leave this interesting issue, and
the construction of the fully back reacted black hole solutions to future work.
One particularly interesting issue is to determine the zero temperature ground state of the system. There are two
natural candidates for such a ground state solution: the new Lif$_4(z\sim 39)$ solution and the non-supersymmetric $AdS_4$ solution.

The supersymmetric $AdS_4\times H^3\times S^4$ solution of
$D=11$ supergravity \cite{Gauntlett:2000ng}
is a specific example of a general class of supersymmetric $AdS_4\times {\cal N}_7$
solutions with magnetic four-form flux, all
describing M5-branes wrapping SLag 3-cycles, that were classified using $G$-structures
in section 9.5 of \cite{Gauntlett:2006ux}. As we discussed in section 6 there is a consistent KK reduction on
any of these ${\cal N}_7$ to minimal gauged supergravity in $D=4$ \cite{Gauntlett:2007ma}.
Given we have shown in this paper that for
the specific example when ${\cal N}_7=H^3\times S^4$ (with suitable twisting and four-form flux)
there is a much bigger
consistent KK reduction, it would be interesting to know whether there is a similarly enlarged
KK truncation for other ${\cal N}_7$.

\subsection*{Acknowledgements}
We would like to thank Guillaume Bossard, Sean Hartnoll, Shamit Kachru, Hermann Nicolai, Tomas Ortin, Maria J. Rodriguez, Julian Sonner, Kelly Stelle, Toby Wiseman
and Daniel Waldram for helpful discussions. AD is supported by an EPSRC Postdoctoral Fellowship. JPG is supported by an EPSRC Senior Fellowship and a Royal Society Wolfson Award and would also like to thank the Perimeter Institute for hospitality.
NK is supported by the National Research Foundation of Korea (NRF)
funded by the Korea government (MEST) by the
grant No. 2009-0085995 and by the grant No. 2005-0049409 through
the Center for Quantum Spacetime (CQUeST) of Sogang University.
OV is supported by an Alexander von Humboldt postdoctoral fellowship and, partially, through the Spanish Government research grant FIS2008-01980. OV also wishes to thank the Galileo Galilei Institute for Theoretical Physics, Florence, where this work was completed, and the INFN for partial support.

\appendix
\section{Consistent KK Truncation formulae}
\subsection{$D=7$ gauged supergravity equations of motion}
We begin by recording the equations of motion for $D=7$ gauged supergravity arising
from \eqref{d7lag}:
\bea \label{h4eq}
&& DS_\3^i  = g T_{ij}\, {* S_\3^j} + \frac{1}{8}  \ep_{i
{j_1}\cdots {j_4}} F_\2^{{j_1} {j_2}}\wedge\, F_\2^{{j_3} {j_4}} \,, \\[12pt]
&& {D\Big(T^{-1}_{ik} T^{-1}_{j\ell} {*F_\2^{ij}}\Big)} =-2g \,
T^{-1}_{i[k} {*DT_{\ell] i}}
-\frac{1}{2g} \,
\ep_{i_1 i_2 i_3 k \ell}\, F_2^{i_1 i_2}\w DS_\3^{i_3}
\nn&&
\qquad \qquad  +\frac{3 }{2g} \delta_{i_1 i_2 k\ell}^{j_1 j_2 j_3 j_4}\, F_\2^{i_1 i_2}\wedge
F_\2^{j_1 j_2}\wedge  F_\2^{j_3 j_4} 
- S_\3^k\wedge S_\3^\ell =0\,,\label{gaugev}\\[12pt]
&& D\Big(\, T^{-1}_{ik} {*D(T_{kj}})\Big)
=2g^2 (2 T_{ik}\, T_{kj} -
T_{kk}\,
T_{ij})\ep_\7 + T^{-1}_{im}\, T^{-1}_{k\ell}\,
{*F_\2^{m\ell}}\wedge F_\2^{kj}
+T_{jk}\, {*S_\3^k} \wedge S_\3^i
  \nn
&&  \qquad   -\frac{1}{5} \delta_{ij}
\Big[2g^2\Big(2T_{ik} T_{ik} -  (T_{ii})^2 \Big) \ep_\7  +
T^{-1}_{nm} T^{-1}_{k\ell}\, {*F_\2^{m\ell}} \wedge F_\2^{kn} +
 T_{k\ell } \, {*S_\3^k} \wedge S_\3^\ell \Big]\,, \label{scalarsv} \\[10pt]
&& \label{7dEinstein}
R_{\mu\nu}=\tfrac{1}{4}T^{-1}_{ij}D_\mu T_{jk} T_{kl}^{-1}D_\nu T_{li}
+\tfrac{1}{4}T^{-1}_{ik}T^{-1}_{jl}F^{ij}_{\mu\rho}F^{kl\rho}_\nu
+\tfrac{1}{4}T_{ij}S^i_{\mu\rho_1\rho_2}S^{j\rho_1\rho_2}_\nu+\tfrac{1}{10}g_{\mu\nu}X\,,
\eea
where
\bea
X&\equiv &-\tfrac{1}{4}T^{-1}_{ik}T^{-1}_{jl}F^{ij}_{\rho_1\rho_2}F^{kl\rho_1\rho_2}
-\tfrac{1}{3}T_{ij}S^i_{\rho_1\rho_2\rho_3}S^{j\rho_1\rho_2\rho_3}+2V\,.
\eea
A typo in \cite{Cvetic:2000ah} has been fixed in (\ref{scalarsv}).

\subsection{Consistency}

We now substitute the KK ansatz (\ref{KKmetricf}), (\ref{KKvectors}), (\ref{KK3formf}) and (\ref{KKscalarsf}) into the equations of motion (\ref{h4eq})--(\ref{7dEinstein}) of $D=7$ maximal gauged supergravity.  To carry out the computation it is helpful to note that
the ansatz implies that
\bea
DT^{ab}&=&-4e^{-4\lambda}d\lambda \delta_{ab}\,,\nn
DT^{a\alpha}&=&g\left[e^{6\lambda}(\cT\theta)_\alpha-e^{-4\lambda}\theta_\alpha\right]\bar e^a\,,\nn
DT^{\alpha\beta}&=&e^{6\lambda}\left[6d\lambda \cT_{\alpha\beta}+D\cT_{\alpha\beta}\right]\,.
\eea
Furthermore
\bea
F_\2^{ab} &=& \tfrac{1}{g}\bar{\cal R}_{ab} -
g \left(\theta^T\theta+\beta^2\right)\bar{e}^a\w\bar{e}^b  - \ep_{abc} \bar{e}^c\w d\beta
\nn
&=&
\tfrac{g}{2}\left[l-2(\theta^T\theta+\beta^2)\right]\bar{e}^a\w\bar{e}^b  - \ep_{abc} \bar{e}^c\w d\beta\,,
\nn
F_\2^{a\alpha}&=& D\theta_\alpha \wedge \bar{e}^a- g  \beta \theta_\alpha \ep_{abc} \bar{e}^b \wedge\bar{e}^c\,,
\nn
F_\2^{\alpha\beta}&=&\ep_{\alpha\beta} F_2\,,
\eea
where $\bar{\cal R}_{ab}$ is the Riemann tensor of $ds^2(\Sigma_3)$ in (\ref{KKmetricf}), we have defined $F_2 = dA_1$,
and
\bea
DS^a_\3 &=& - \bar{e}^a \w (dB_2-g \theta_\alpha h_\alpha)
+\ep_{abc} \bar{e}^b \w\bar{e}^c \w ( dC_1 - g \beta B_2)\,,
\nn
DS_\3^{\alpha}
&=&
-g \mathrm{vol}(\Sigma_3) \w ( D\chi_\alpha + 6 \theta_\alpha C_1 )
+ Dh_\alpha\,.
\eea
Using the obvious orthonormal frame in $D=7$ we can calculate the components of the $D=7$ Ricci-tensor
and find:
\bea
R_{mn}&=&e^{6\phi}\left[R^{(4)}_{mn}+3\nabla^2\phi\eta_{mn}-30\nabla_m\phi\nabla_n\phi \right]\,,\nn
R_{am}&=&0\,,\nn
R_{ab}&=&e^{6\phi}\left[-2\nabla^2\phi+lg^2e^{-10\phi}\right]\delta_{ab} \, ,
\eea
$m=0,1,2,3$. We also find
\bea
-e^{-6\phi}X&=&
3e^{8\lambda-4\phi}(\nabla\beta)^2
+3e^{-2\lambda-4\phi}(D_m\theta^T\cT^{-1}D^m\theta)
+\tfrac{1}{2}e^{-12\lambda+6\phi}F_{mn}F^{mn}\nn
&+&\frac{3g^2}{4}\left[l-2(\theta^T\theta+\beta^2)\right]^2e^{8\lambda-14\phi}
+12g^2\beta^2e^{-2\lambda-14\phi}(\theta^T\cT^{-1}\theta)\nn
&+&3e^{-4\lambda+2\phi}B_{mn}B^{mn}
+24e^{-4\lambda-8\phi}C_m C^m
+\tfrac{1}{3}e^{6\lambda+12\phi}(h^T_{p_1p_2p_3}\cT h^{p_1p_2p_3})\nn
&+&2g^2 e^{6\lambda-18\phi}(\chi^T\cT\chi)\nn
&+&g^2e^{-6\phi}\Big[
3e^{-8\lambda}+e^{12\lambda}[ (\Tr\cT)^2-2\Tr(\cT\cT)]+6e^{2\lambda}\Tr\cT\Big]\,.
\eea

Proceeding, we now find that Eq (\ref{h4eq}) gives:
\bea\label{firstsetone}
&& dB_2  -g (\theta^T h_3) +2 e^{-4\lambda-8\phi} g *C_1    -  d\beta \wedge F_2 =0\,, \\[10pt]
&& dC_1 -g \beta B_2 - \tfrac12  e^{-4\lambda+2\phi} g *B_2
-\frac{g}{4}\left[l-2(\theta^T\theta+\beta^2)\right]F_2 -\frac{1}{2} \epsilon_{\alpha \beta} D \theta^\alpha \wedge   D \theta^\beta = 0
\label{firstsettwo} \nn[10pt]
\eea
and
\bea
&& D h_3^\alpha - e^{6\lambda-18\phi} g^2 (\cT\chi)_\alpha \textrm{vol}_4 = 0\,,\label{firstsetthree} \\[10pt]
&& D\chi_\alpha +6 \theta_\alpha C_1
+e^{6\lambda+12\phi}  *(\cT h_3)_\alpha
-6\epsilon_{\alpha \beta} \theta^\beta \beta d\beta
+\tfrac{3}{2}\left[l-2(\theta^T\theta+\beta^2)\right]
\epsilon_{\alpha \beta} D\theta^\beta=0\,,\label{firstsetfour}
\nn
\end{eqnarray}
where $*$ and $\textrm{vol}_4$ are the Hodge dual and volume form corresponding to the four-dimensional metric $ds^2_4$ in (\ref{KKmetricf}). Observe that (with $g\ne 0$) these equations imply that
\begin{eqnarray}
&& d(e^{-4\lambda +2\phi} *B_2) + 2g \beta \theta^T h_3 - 4g
e^{-4\lambda  -8\phi} \beta *C_1 + 2 B_2 \wedge d\beta =0\,, \label{secondsetone}  \\[10pt]
&& d ( e^{-4\lambda -8 \phi} *C_1 ) -\tfrac12 D\theta^T \wedge h_3
-\tfrac12 g^2 e^{6\lambda -18 \phi} (\theta^T \cT \chi)  \textrm{vol}_4 =0\,, \label{secondsettwo}
\\[10pt]
&& D ( e^{6\lambda+12 \phi} \cT *h_3 ) +3g e^{-4\lambda+2 \phi} \theta
*B_2 - 6 C_1 \wedge D\theta + 6g \beta \theta B_2 + g \epsilon \chi F_2  =0\,.\nn\label{secondsetthree}
\end{eqnarray}

Next we find that Eq (\ref{gaugev}) gives:
\begin{eqnarray}
&& d(e^{-4\phi +8\lambda} * d \beta) + g^2 \beta  \Big[  4 e^{-14\phi -2\lambda} ( \theta^T \cT^{-1} \theta) -  e^{-14 \phi +8\lambda} (l-2\beta^2 -2 \theta^T \theta)  \nonumber  \\
&& \qquad -2 e^{-18 \phi +6\lambda}  ( \theta^T \epsilon \cT \chi ) \Big] \textrm{vol}_4  +e^{-4\lambda+2\phi} F_2 \wedge  *B_2 +  B_2 \wedge B_2 = 0 \; , \label{betaeom} \\[20pt]
%
%
&& D(e^{-4\phi -2\lambda}  \cT^{-1} *D \theta) + g^2 \Big[ 4e^{-14\phi -2\lambda}\beta^2 \cT^{-1} \theta - e^{-14\phi +8\lambda}  (l-2\beta^2 -2 \theta^T \theta) \theta \nonumber \\
&& \qquad + e^{-10\phi} ( e^{10\lambda} \cT \theta - 2\theta + e^{-10\lambda} \cT^{-1} \theta) +\ft12  e^{-18\phi +6\lambda} (l-2\beta^2 -2 \theta^T \theta) \epsilon  \cT \chi  \Big] \textrm{vol}_4 \nn
&& \qquad +4  e^{-4\lambda-8\phi}  *C_1    \wedge \epsilon D \theta   + 2  C_1 \wedge h_3 = 0 \; , \label{thetaeom} \\[20pt]
&& d (e^{6\phi-12\lambda} *F_2 )
+3g  e^{-4\phi-2\lambda} ( \theta^T \epsilon \cT^{-1} *D\theta)
+g \Tr(\epsilon \cT^{-1}  *D\cT ) - g (  \chi^T \epsilon h_3) \nonumber \\
&& \qquad - 3g e^{-4\lambda-8\phi}(l-2\beta^2 -2 \theta^T \theta) *C_1
+3  e^{-4\lambda+2\phi} d\beta \wedge  *B_2
   = 0  \label{F2eom}
\end{eqnarray}
and we have used \eqref{firstsetone}--\eqref{firstsetthree} to simplify the expressions.

We next turn to  (\ref{scalarsv}).
From the $\left(ab \right)$ components we obtain
\begin{eqnarray} \label{lambda_eom2}
&& d*d\lambda -\tfrac{1}{5}e^{8\lambda-4\phi}d \beta \wedge *d\beta
+\tfrac{1}{20}e^{-2\lambda-4\phi} D \theta^T \wedge * \cT^{-1} D\theta
-\tfrac{1}{20}e^{6\lambda+12\phi} h^T_3 \wedge * \cT h_3  \nonumber \\
&& \qquad -\tfrac{1}{10}e^{-12\lambda+6\phi}F_2 \wedge * F_2
-\tfrac{1}{10}e^{-4\lambda+2\phi}B_2 \wedge *B_2 +\ft{2}{5}
e^{-4\lambda-8\phi}C_1 \wedge *C_1
 \nonumber \\
&& \qquad +g^2 \Big\{  \ft{1}{20}
e^{8\lambda-14\phi} (l -2 \beta^2 - 2 \theta^T \theta)^2   \nonumber \\
&& \qquad \qquad \   +\ft{1}{10} e^{-6\phi} \left[
2e^{-8\lambda} -e^{12\lambda}[ (\Tr\cT)^2-2\Tr(\cT{\cal
T})] -e^{2\lambda}\Tr\cT \right]  \nonumber \\
&& \qquad \qquad \ +\tfrac{1}{4} e^{-10\phi}
\left[e^{10\lambda}(\theta^T\cT\theta)- e^{-10\lambda}
(\theta^T\cT^{-1}\theta) \right] \nonumber \\
&& \qquad \qquad \ -\ft{1}{5} e^{-2\lambda-14\phi} \beta^2
(\theta^T\cT^{-1}\theta) +  \ft{1}{20}
e^{6\lambda-18\phi}(\chi^T\cT\chi) \Big\} \textrm{vol}_4 = 0\,.
\end{eqnarray}
From the $\left(\alpha\beta \right)$ components of \eqref{scalarsv}, and also using
(\ref{lambda_eom2}),
we obtain\begin{eqnarray}\label{cos_scalars2}
&& D\left(\cT^{-1}_{\alpha\gamma} * D\left(\cT^\gamma{}_\beta \right) \right) +\ft32 e^{-4\phi-2\lambda} \left(2\cT^{-1}_{\alpha \delta} \delta_{\beta\gamma} - \cT^{-1}_{\gamma \delta} \delta_{\alpha\beta} \right) D\theta^\gamma \wedge *D\theta^\delta \nonumber \\
&& \qquad -\ft12 e^{12\phi+6\lambda} \left(2\cT_{\beta \delta} \delta_{\alpha\gamma} - \cT_{\gamma \delta} \delta_{\alpha\beta} \right) h_3^\gamma \wedge * h_3^\delta \nonumber \\
&& \qquad +g^2 \Big\{ e^{-6\phi} \left[ 4 e^{12\lambda} (\cT^2)_{\alpha\beta}  -2e^{12\lambda} \cT_{\alpha\beta} \Tr \cT
 - e^{12\lambda} \delta_{\alpha\beta} \left( 2\Tr (\cT\cT) - (\Tr \cT)^2 \right)  \right. \nonumber \\
&& \qquad \qquad \qquad \quad \left. -6 e^{2\lambda }\cT_{\alpha\beta} + 3 e^{2\lambda} \delta_{\alpha\beta} \Tr \cT \right] \nonumber \\
&& \qquad \qquad +\ft32 e^{-10\phi} \left[ e^{10\lambda} \left(2\cT_{\beta \delta} \delta_{\alpha\gamma} - \cT_{\gamma \delta} \delta_{\alpha\beta} \right) -e^{-10\lambda} \left(2\cT^{-1}_{\alpha \delta} \delta_{\beta\gamma} - \cT^{-1}_{\gamma \delta} \delta_{\alpha\beta} \right) \right] \theta^\gamma \theta^\delta \nonumber \\
&& \qquad \qquad -6 e^{-2\lambda-14\phi} \beta^2 \left(2\cT^{-1}_{\alpha \delta} \delta_{\beta\gamma} - \cT^{-1}_{\gamma \delta} \delta_{\alpha\beta} \right) \theta^\gamma \theta^\delta \nonumber \\
&& \qquad \qquad +\ft{1}{2} e^{6\lambda-18\phi}\left(2\cT_{\beta \delta} \delta_{\alpha\gamma} - \cT_{\gamma \delta} \delta_{\alpha\beta} \right)\chi^\gamma \chi^\delta  \Big\} \ \textrm{vol}_4 = 0\,.\label{enneom}
\end{eqnarray}
The mixed $\left(a\alpha\right)$ components of \eqref{scalarsv} are trivially satisified.

Finally, we consider the Einstein equations \eqref{7dEinstein}.
The $ab$ components of \eqref{7dEinstein} give:
\begin{eqnarray} \label{4dbreathing}
&& d*d\phi +\tfrac{1}{10}e^{8\lambda-4\phi}d \beta \wedge *d\beta
+\tfrac{1}{10}e^{-2\lambda-4\phi} D \theta^T \wedge * \cT^{-1} D\theta
-\tfrac{1}{10}e^{6\lambda+12\phi} h^T_3 \wedge * \cT h_3  \nonumber \\
&& \qquad +\tfrac{1}{20}e^{-12\lambda+6\phi}F_2 \wedge * F_2
+\tfrac{1}{20}e^{-4\lambda+2\phi}B_2 \wedge *B_2 +\ft{4}{5}
e^{-4\lambda-8\phi}C_1 \wedge *C_1
 \nonumber \\
&& \qquad +g^2 \Big\{ \ft12 l e^{-10\phi}   -\ft{7}{80}
e^{8\lambda-14\phi} (l -2 \beta^2 - 2 \theta^T \theta)^2   \nonumber \\
&& \qquad \qquad \   +\ft{1}{20} e^{-6\phi} \left[
3e^{-8\lambda}+e^{12\lambda}[ (\Tr\cT)^2-2\Tr(\cT\cT)]+6e^{2\lambda}\Tr\cT \right]  \nonumber \\
&& \qquad \qquad \ -\tfrac{1}{4} e^{-10\phi}
\left[e^{10\lambda}(\theta^T\cT\theta)-2\theta^T\theta + e^{-10\lambda}
(\theta^T\cT^{-1}\theta) \right] \nonumber \\
&& \qquad \qquad \ -\ft{7}{5} e^{-2\lambda-14\phi} \beta^2
(\theta^T\cT^{-1}\theta) -  \ft{3}{20}
e^{6\lambda-18\phi}(\chi^T\cT\chi) \Big\} \textrm{vol}_4 = 0\,.
\end{eqnarray}
The $mn$ components of \eqref{7dEinstein}, after using
\eqref{4dbreathing}, lead to the following $D=4$ Einstein equations
(using an orthonormal frame associated with $ds^2_4$) :
\begin{eqnarray} \label{4dEinstein}
R^{(4)}_{mn} &=& 30 \nabla_m \phi \nabla_n \phi + 30 \nabla_m \lambda \nabla_n \lambda  + \ft14 \Tr(\cT^{-1}D_m\cT\cT^{-1}D_n\cT) \nonumber \\
&& +\ft32 e^{8\lambda-4\phi} \nabla_m \beta \nabla_n \beta +\tfrac{3}{2}e^{-2\lambda-4\phi}
\Tr(D_m\theta \cT^{-1} D_n\theta) \nonumber \\
&& +\tfrac{1}{2}e^{-12\lambda+6\phi} \left( F_{mp}F_n{}^p -\ft14 \eta_{mn} F_{pq}F^{pq} \right) +\tfrac{3}{2}e^{-4\lambda+2\phi} \left( B_{mp}B_n{}^p -\ft14 \eta_{mn} B_{pq}B^{pq} \right) \nonumber \\
&& +6e^{-4\lambda-8\phi}C_m C_n +\tfrac{1}{4}e^{6\lambda+12\phi} \left( h^T_{mp_1p_2}\cT h_n{}^{p_1p_2} -\ft13 \eta_{mn} h^T_{p_1p_2p_3}\cT h^{p_1p_2p_3} \right) \nonumber  \\
&& -g^2 \eta_{mn} \Big\{ \ft32 l e^{-10\phi}  -\ft{3}{16} e^{8\lambda-14\phi} (l -2 \beta^2 - 2 \theta^T \theta)^2 \nonumber \\
&& \qquad \qquad   +\ft14  e^{-6\phi} \left[ 3e^{-8\lambda}+e^{12\lambda}[ (\Tr\cT)^2-2\Tr(\cT\cT)]+6e^{2\lambda}\Tr\cT \right]  \nonumber \\
&& \qquad \qquad -\tfrac{3}{4} e^{-10\phi}\left[e^{10\lambda}(\theta^T\cT\theta)-2\theta^T\theta+e^{-10\lambda}
(\theta^T\cT^{-1}\theta)\right] \nonumber \\
&& \qquad \qquad -3e^{-2\lambda-14\phi} \beta^2 (\theta^T\cT^{-1}\theta) -  \ft{1}{4} e^{6\lambda-18\phi}(\chi^T\cT\chi) \Big\}\,.
\end{eqnarray}
The $ma$ components of \eqref{7dEinstein} are trivially satisfied.

We have thus demonstrated that the KK truncation ansatz is consistent. Any solution of the
$D=4$ equations of motion given in \eqref{firstsetone}-\eqref{4dEinstein}
gives rise to a solution of
$D=11$ supergravity after uplifting first to $D=7$ via
(\ref{KKmetricf}), (\ref{KKvectors}), (\ref{KK3formf}) and (\ref{KKscalarsf}), and then to $D=11$ through (\ref{metel}), (\ref{4form}).

We end this appendix by noting that
these $D=4$ equations of motion remain inert under the scaling
\bea\label{scalingsymmetry}
g_{mn}&\to& L^2 g_{mn}\,,\nn
F_{mn}&\to& L F_{mn}\,,\nn
B_{mn}&\to& L B_{mn}\,,\nn
C_{m}&\to& C_{m}\,,\nn
h_{mnp}&\to& L^2 h_{mnp}\,,\nn
g&\to& L^{-1} g\,.
\eea

\section{Simplified $B_2$, $C_1$ system}
\label{lastapp}
Consider the following Lagrangian in flat space
\be\label{zlag}
{\cal L}=-\frac{1}{2}B_2\wedge *B_2+\frac{m^2}{2}C_1\wedge *C_1+C_1\wedge dB_2\,.
\ee
This describes, somewhat unconventionally, a massive vector field.
The equations of motion are
\bea
dC_1&=&*B_2\,,\label{zip}\\
m^2*C_1+dB_2&=&0\,.\label{zop}
\eea
We can solve the $B_2$ equation of motion \eqref{zip} for $B_2$, $B_2=-*dC_1$, and then substitute into \eqref{zop}  to get
\be\label{ink}
d*dC_1-m^2*C_1=0\,,
\ee
which is the usual equation for a massive spin 1 field. Note also that we can substitute into the Lagrangian \eqref{zlag}
to get
\be
{\cal L}=-\frac{1}{2}dC_1\wedge *dC_1+\frac{m^2}{2}C_1\wedge *C_1\,,
\ee
which leads to the same equation of motion \eqref{ink}.

Alternatively, from \eqref{zip} we deduce that
$d*B_2=0$ which we can solve by writing
\be\label{pink}
B_2=-*d\tilde B_1
\ee
and we observe that $\tilde B_1$ is only defined up to a gauge transformation $\tilde B_1\to\tilde B_1+d\Lambda$.
Equation \eqref{zip} can then be written $dC_1=d\tilde B_1$, which is solved via
\be\label{ponk}
C_1=\tilde B_1 +db
\ee
and notice that this maintains the gauge invariance provided that $b\to b-\Lambda$.
In terms of these variables \eqref{zop} can be written as
\be
d*d\tilde B_1-m^2*(\tilde B_1+db)=0\,.
\ee
Note that this comes
from a Lagrangian which can be obtained by substituting \eqref{pink},\eqref{ponk} into \eqref{zlag}, namely
\be
{\cal L}=-\frac{1}{2}\tilde H_2\wedge *\tilde H_2+\frac{m^2}{2}(\tilde B_1+db)\wedge * (\tilde B_1+db)\,,
\ee
where $\tilde H_2=d\tilde B_1$. This is the standard Stuckelberg Lagrangian.

\section{The hypermultiplet moduli space} \label{hyperMS}

Here we fix our conventions for the Lie algebra $\mathfrak{g}_2$ of the group $G_2$, and give details of the construction of the hypermultiplet moduli space $G_{2(2)}/SO(4)$. This is the eight-dimensional quaternionic-K\"ahler, symmetric space associated to the split, maximally noncompact real form $\mathfrak{g}_{2(2)}$. It is this real form we will be referring to when we write $\mathfrak{g}_2$ below.

\subsection{$G_2$ conventions}

We find it convenient to choose the following set of positive roots for $\mathfrak{g}_2$,
\begin{eqnarray}
\mathrm{
\begin{tabular}{ll}
$\vec\alpha_1 =  (0,2)$,& \quad
$\vec\alpha_2 = (-\frac{1}{\sqrt 3},-1)$,\\
$\vec\alpha_3 =  (-\frac{1}{\sqrt 3},1)=\vec\alpha_1+\vec\alpha_2$,& \quad
$\vec\alpha_4 =  (-\frac{2}{\sqrt 3},0)=\vec\alpha_1+2\vec\alpha_2$, \\
$\vec\alpha_5 = (-\sqrt 3,-1)=\vec\alpha_1+3\vec\alpha_2$,& \quad
$\vec\alpha_6 = (-\sqrt 3,1)=2\vec\alpha_1+3\vec\alpha_2$,
\end{tabular}
}
\label{posroots}
\end{eqnarray}
with $\vec\alpha_1$, $\vec\alpha_2$ as the simple roots\footnote{The convenience of this choice can be see from the Lagrangian (\ref{HMpara}): it is this set of roots that governs the couplings of the dilatons to the axion kinetic terms in the hypermultiplet sector of our theory.}. We collectively denote the two Cartan generators $\textsf{H}_1$, $\textsf{H}_2$ as $\vec{\textsf{H}}$, and the six positive and six negative root generators as $\textsf{E}_{i} \equiv \textsf{E}_{\vec \alpha_i}$ and $\textsf{F}_{i} \equiv \textsf{E}_{-\vec \alpha_i}$, $i=1, \ldots, 6$, respectively. The canonical commutation relations read
\begin{eqnarray} \label{CWcomm1}
&& [\textsf{H}_1 , \textsf{H}_2] =0\,,  \qquad \quad  [\textsf{E}_{\vec\alpha} , \textsf{E}_{-\vec\alpha}] = \tfrac{1}{2} \vec \alpha \cdot \vec{\textsf{H}}\,,
 \nonumber \\
&& [ \vec{\textsf{H}}, \textsf{E}_{\vec \alpha}] = \vec\alpha \textsf{E}_{\vec\alpha}\,,  \qquad
[\textsf{E}_{\vec\alpha} , \textsf{E}_{\vec\beta}] = N_{\vec\alpha , \vec\beta} \textsf{E}_{\vec \alpha + \vec \beta}\,,
\end{eqnarray}
where the non-vanishing structure constants $N_{\vec\alpha , \vec\beta}$ are given by
\begin{eqnarray} \label{CWcomm2}
 N_{\vec \alpha_1 , \vec \alpha_2} = N_{\vec \alpha_1 , \vec \alpha_5} = -N_{\vec \alpha_2 , \vec \alpha_4} = -N_{\vec \alpha_3 , \vec \alpha_4} = 1 \; , \quad N_{\vec \alpha_2 , \vec \alpha_3} = -\tfrac{2}{\sqrt{3}} \, ,
\end{eqnarray}
together with the relations
\begin{eqnarray} \label{CWcomm3}
N_{\vec \alpha , \vec \beta} =  -N_{\vec \beta  , \vec \alpha} = - N_{-\vec \alpha , -\vec \beta}
= N_{\vec \beta  , -\vec \alpha- \vec \beta} = N_{-\vec \alpha- \vec \beta  , \vec \alpha}\,.
\end{eqnarray}

These $\mathfrak{g}_2$ commutation relations are all that is needed to compute all the quantities we are interested in. For calculational purposes, however, it proves helpful to have an explicit relation of the generators of $\mathfrak{g}_2$. Calling $E_{ij}$ the $7\times 7$ matrix with 1 in the $i$-th row and $j$-th column and 0 elsewhere, an explicit realisation of the $\mathfrak{g}_2$ generators in the fundamental representation is given by
\begin{center}
\begin{tabular}{ll}
$\textsf{H}_1 = \tfrac{1}{\sqrt{3}} \left(E_{11} -E_{22} +2 E_{33}-2E_{55}+E_{66}-E_{77} \right)$ &
$\textsf{H}_2 = E_{11} +E_{22} - E_{66} -E_{77}$ \\
$\textsf{E}_1 = -2E_{16} - 2E_{27}$ &
$\textsf{F}_1 = -\tfrac{1}{2} \left( E_{61} +E_{72} \right)$ \\
$\textsf{E}_2 = \tfrac{1}{2\sqrt{3}} \left(2E_{41} -E_{52} - E_{63} +2E_{74} \right)$  &
$\textsf{F}_2 = \tfrac{2}{\sqrt{3}} \left(E_{14} -E_{25} - E_{36} +E_{47} \right)$ \\
$\textsf{E}_3 = \tfrac{1}{\sqrt{3}} \left(E_{13} -2E_{24} +2 E_{46} -E_{57} \right)$ &
$\textsf{F}_3 = \tfrac{1}{\sqrt{3}} \left(E_{31} -E_{42} +E_{64} -E_{75} \right)$ \\
$\textsf{E}_4 = -\tfrac{1}{\sqrt{3}} \left(E_{21} +E_{43} +E_{54} +E_{76} \right)$ &
$\textsf{F}_4 = -\tfrac{1}{\sqrt{3}} \left(E_{12} +2E_{34} +2E_{45} +E_{67} \right)$ \\
$\textsf{E}_5 = \tfrac{1}{2} \left( -E_{51} +E_{73} \right)$ &
$\textsf{F}_5 = -2 E_{15} +2 E_{37}$ \\
$\textsf{E}_6 = -E_{23} -E_{56}$ &
$\textsf{F}_6 = -E_{32} -E_{65}$
\end{tabular}
\\[-20pt]
\begin{eqnarray}
\end{eqnarray}
\end{center}
It can be checked that this set of 7-dimensional matrices satisfy the commutation relations (\ref{CWcomm1}) with (\ref{CWcomm2}), (\ref{CWcomm3}).

A couple of operations that are needed for our analysis are the Cartan involution $\tau$ and the generalised transpose $\sharp$. The former is the Lie algebra automorphism defined through its action on the Cartan-Weyl basis as
\begin{equation}
 \tau(\vec{\textsf{H}}) = - \vec{\textsf{H}} \ , \quad \tau(\textsf{E}_i) = - \textsf{F}_i \, \quad \tau(\textsf{F}_i) = - \textsf{E}_i \ , \quad i=1,\ldots, 6.
\end{equation}
The $\tau$-invariant subalgebra, spanned by the generators $\textsf{K}_i =\textsf{E}_i-\textsf{F}_i$, $i=1,\ldots, 6$ is the maximal compact subalgebra $\mathfrak{so}(4)$. Indeed the combinations $\textsf{J}_x$, $\textsf{L}_x$, $x=1,2,3$,
\begin{eqnarray}
\mathrm{
\begin{tabular}{ll}
$\textsf{J}_1 = \tfrac14 \left( \textsf{K}_5 -\sqrt{3} \textsf{K}_3 \right), $ \qquad \qquad&
$\textsf{L}_1 = \tfrac14 \left( 3\textsf{K}_5 +\sqrt{3} \textsf{K}_3 \right)\,, $ \\
$\textsf{J}_2 = \tfrac14 \left( \textsf{K}_6 +\sqrt{3} \textsf{K}_2 \right), $ &
$\textsf{L}_2 = \tfrac14 \left( 3\textsf{K}_6 -\sqrt{3} \textsf{K}_2 \right)\,,$ \\
$\textsf{J}_3 = \tfrac14 \left( \textsf{K}_1 -\sqrt{3} \textsf{K}_4 \right), $ &
$\textsf{L}_3 = \tfrac14 \left( 3\textsf{K}_1 +\sqrt{3} \textsf{K}_4 \right)\,, $ \\
\end{tabular}
}
\end{eqnarray}
can be checked to satisfy the canonical $SO(4) \approx SO(3)\times SO(3)$ commutation relations
\begin{equation} \label{SO4commrel}
 [\textsf{J}_x , \textsf{J}_y ] = \epsilon_{xyz} \textsf{J}_z \, , \quad
[\textsf{J}_x , \textsf{L}_y] =0 \, ,\quad
[\textsf{L}_x , \textsf{L}_y ] =  \epsilon_{xyz} \textsf{L}_z \ .
\end{equation}

The generalised transpose $\sharp$ can be defined, at the Lie algebra level, as $\sharp = -\tau$:
\begin{equation}
 \sharp(\vec{\textsf{H}}) =  \vec{\textsf{H}} \ , \quad \sharp(\textsf{E}_i) =  \textsf{F}_i \, \quad \sharp(\textsf{F}_i) =  \textsf{E}_i \ , \quad i=1,\ldots, 6.
\end{equation}
We are also interested in the action of the generalised transpose at the group level, which can be defined through the exponential map: for $\textsf{X} \in \mathfrak{g}_2$ with group element $g=e^{\textsf{X}} \in G_2$, we can define $g^\sharp=e^{\sharp(\textsf{X})}$. Relevant properties of $\sharp$ are $(g^\sharp)^\sharp = g$, which follows from its definition at the Lie algebra level, and  $(g_1 g_2)^\sharp = g_2^\sharp g_1^\sharp$ for all $g_1 ,   g_2  \in G_2$, which can be shown with the help of the Baker-Campbell-Hausdorff formula.

\subsection{The $G_{2(2)}/SO(4)$ coset space}

The Iwasawa decomposition (see e.g. \cite{popesnotes}) of $\mathfrak{g}_{2}$ can be invoked to construct a coset representative of the maximally non-compact space $G_{2(2)}/SO(4)$ via the exponentiation of the Borel subalgebra of Cartan and positive root generators:
\begin{eqnarray} \label{coset}
{\cal V} = e^{\frac{1}{2} \vec\varphi \cdot \vec{\textsf{H}} } e^{\zz \textsf{E}_1} e^{\sqrt{3}(-\theta_1 \textsf{E}_2 + \theta_2 \textsf{E}_3)} e^{\xi_2 \textsf{E}_6} e^{2\sqrt{3}  \ab \textsf{E}_4 -\xi_1 \textsf{E}_5 } \, ,
\end{eqnarray}
where
\begin{equation} \label{coordsG2SO4}
 q^u \equiv (\varphi_1, \varphi_2, \zz, \theta_1, \theta_2, \ab, \xi_1, \xi_2) \ , \quad u=1, \ldots, 8 \, ,
\end{equation}
are coordinates on $G_{2(2)}/SO(4)$, and the numerical factors in the exponentials have been choosen to make contact with the main text. In (\ref{coset}) we have defined $\vec\varphi = (\varphi_1, \varphi_2)$, and have used a dot to denote the usual Euclidean scalar product.

A metric on $G_{2(2)}/SO(4)$ can be constructed as follows. First introduce the right-invariant one-forms ${\cal F}_1^i$, $i=1, \ldots , 6$, defined as
\begin{eqnarray} \label{oneformFSg=0}
{\cal F}_1^1 &=& d \zz \; , \nonumber \\[5pt]
{\cal F}_1^2 &=& \sqrt{3} \  d \theta^1 \; , \nonumber \\[5pt]
{\cal F}_1^3 &=&  \sqrt{3} \ (d\theta^2 - \zz d \theta^1) \; ,
\nonumber \\[5pt]
{\cal F}_1^4 &=& 2\sqrt{3} \ G_1 =2 \sqrt{3} \left(d\ab + \tfrac12(
\theta^1 d \theta^2 -\theta^2 d \theta^1 )\right) \; ,  \\[5pt]
{\cal F}_1^5 &=& F^1_1 =  d \xi^1 -6\theta^1  d\ab - \theta^1 (\theta^1
d\theta^2 -\theta^2 d\theta^1 )  \; , \nonumber \\[5pt]
{\cal F}_1^6 &=& F^2_1  - \zz F^1_1 = \nonumber \\
 &=& d \xi^2 -6\theta^2 d\ab - \theta^2 (\theta^1 d\theta^2 -\theta^2
d\theta^1 ) -\zz \left( d\xi^1 -6\theta^1  d\ab  - \theta^1 (\theta^1
d\theta^2 -\theta^2 d\theta^1 ) \right) \; .  \nonumber
\end{eqnarray}
The right-invariant Maurer-Cartan form associated to the coset representative (\ref{coset}) takes values in the Borel subalgebra. Using the one-forms (\ref{oneformFSg=0}), it can be written as
\begin{eqnarray}
 d{\cal V} \ {\cal V}^{-1} & = &  \tfrac{1}{2} d \vec \varphi \cdot \vec{\textsf{H}} + e^{ \frac{1}{2} \vec{\alpha}_1  \cdot \vec{\varphi}} {\cal F}_1^1 \ \textsf{E}_1 - e^{ \frac{1}{2} \vec{\alpha}_2  \cdot \vec{\varphi}} {\cal F}_1^2 \ \textsf{E}_2 + e^{ \frac{1}{2} \vec{\alpha}_3  \cdot \vec{\varphi}} {\cal F}_1^3 \ \textsf{E}_3 + e^{ \frac{1}{2} \vec{\alpha}_4  \cdot \vec{\varphi}} {\cal F}_1^4 \ \textsf{E}_4 \nonumber \\
&& - e^{ \frac{1}{2} \vec{\alpha}_5  \cdot \vec{\varphi}} {\cal F}_1^5 \ \textsf{E}_5 + e^{ \frac{1}{2} \vec{\alpha}_6  \cdot \vec{\varphi}} {\cal F}_1^6 \ \textsf{E}_6\,.
\end{eqnarray}
%
%
%
Next, introduce the $\mathfrak{g}_2$--valued one-form
\begin{eqnarray}
 P= \tfrac{1}{2}\left( d{\cal V} \ {\cal V}^{-1} + \left( d{\cal V} \ {\cal V}^{-1} \right)^\sharp \right) \; ,
\end{eqnarray}
and the quadratic form
\begin{eqnarray}  \label{hypersmatrix}
 {\cal M} ={\cal V}^ \sharp {\cal V} \; ,
\end{eqnarray}
which are related by
\begin{eqnarray}
 \tfrac{1}{2}  {\cal M}^{-1} d{\cal M}  = {\cal V}^{-1} P  {\cal V} \; .
\end{eqnarray}
A globally $G_2$ right-invariant and locally $SO(4)$ left-invariant metric on the coset space is finally given by either of the two equivalent expressions
\begin{eqnarray}
h_{uv} dq^u dq^v &=& \tfrac14 \Tr \left( P P \right) \nonumber \\
 &=& \tfrac{1}{16} \Tr \left( {\cal M}^{-1} d {\cal M}  {\cal M}^{-1}  d {\cal M} \right) \; .
 \end{eqnarray}
In terms of the forms (\ref{oneformFSg=0}), this is equivalent to
\begin{eqnarray}
h_{uv} dq^u dq^v &=& \tfrac14 d \vec\varphi \cdot d \vec\varphi + \tfrac{1}{4} \sum_{i=1}^{6}  e^{  \vec{\alpha}_i  \cdot \vec{\varphi}} ({\cal F}_1^i)^2  \; ,
 \end{eqnarray}
and, in terms of the explicit coordinates (\ref{coordsG2SO4}) and the one-forms $G_1$, $F_1^\alpha$, this is\footnote{We find agreement with \cite{Compere:2009zh}, up to an overall factor of $1/2$, with the identifications $\chi_1= \zz$, $\chi_2=\sqrt{3} \theta^1$,
$\chi_3=\sqrt{3} \theta^2$, $\chi_4= 2\sqrt{3} \ab$, $\chi_5=\xi^1$,
$\chi_6=\xi^2$.}
\begin{eqnarray} \label{HMg=0metric}
h_{uv} dq^u dq^v &=& \tfrac14 (d \varphi_1)^2
+\tfrac14 (d \varphi_2)^2  + \tfrac14 e^{2\varphi_2} (d \zz )^2
+\tfrac{3}{4}e^{-\frac{1}{\sqrt{3}}\varphi_1-\varphi_2} (d \theta^1)^2 \nonumber \\
&&
+\tfrac{3}{4}e^{-\frac{1}{\sqrt{3}}\varphi_1+\varphi_2} (d\theta^2
- \zz d \theta^1)^2   +3 e^{-\frac{2}{\sqrt{3}}
\varphi_1} (G_1)^2 +\tfrac{1}{4}e^{-\sqrt{3}\varphi_1-\varphi_2} (F^1_1)^2
\nonumber \\
&&  +\tfrac{1}{4}e^{-\sqrt{3}\varphi_1+\varphi_2} (F^2_1 - \zz F^1_1)^2 \ .
\end{eqnarray}

The Riemannian manifold $G_{2(2)}/SO(4)$ equipped with this metric is a quaternionic-K\"ahler space and, therefore, Einstein. Indeed, the Ricci tensor of (\ref{HMg=0metric}) can be checked to give $(-8)$ times the metric (\ref{HMg=0metric}) itself. Being further a symmetric space, its holonomy group coincides with the isotropy group $SO(4)$, which should be thought of as a subgroup of the holonomy group $Sp(1) \times Sp(2)$ of a generic eight-dimensional quaternionic-K\"ahler space. The
$Sp(1)$ factor of the holonomy is related to the existence of a triplet of complex structures $J^x$, $x=1,2,3$, satisfying the quaternion algebra
\begin{equation} \label{quaternion}
 J^x J^y =-\delta^{xy} +\epsilon^{xyz} J^z .
\end{equation}

We now explicitly elucidate the $SO(4)$ holonomy of our quaternionic K\"ahler manifold $G_{2(2)}/SO(4)$, and in particular the
canonical $Sp(1)$ factor giving rise to \eqref{quaternion}. To this end we first introduce the obvious orthonormal frame $e^{\bar u}$, where here $\bar u = 1, \ldots, 8$, are tangent space indices, for the metric (\ref{HMg=0metric}):
\begin{eqnarray} \label{frameG2SO4}
 e^1 = \tfrac{1}{2} d\varphi_1 \; , \quad  e^2 = \tfrac{1}{2} d\varphi_2 \; , \quad
e^{i+2} = \tfrac{1}{2} e^{\tfrac{1}{2} \vec{\alpha}_i  \cdot \vec{\varphi}} {\cal F}_1^i , \quad i=1,\ldots, 6 .
\end{eqnarray}
The corresponding spin connection $w$, satisfying $de^{\bar u} + w^{\bar u}{}_{\bar v} \wedge e^{\bar v} =0 $,
is a one-form valued in the Lie-algebra $\mathfrak{so}(8)$, the holonomy algebra for a generic orientable eight-dimensional manifold.
Introducing the $\mathfrak{so}(8)$ generators $\textsf{M}_{\bar u \bar v} = E_{\bar u \bar v}-E_{\bar v \bar u}$, where $E_{\bar u \bar v}$ is here the $8 \times 8$ matrix with 1 in the $(\bar u,\bar v)$ position and 0 elsewhere, we find that $w$ can be written as
\begin{eqnarray} \label{spincon1}
w =  e^3 \textsf{M}_3 +e^4 \textsf{N}_2 -e^5 \textsf{N}_1 +e^6\textsf{N}_3 +
 e^7 \textsf{M}_1 + e^8 \textsf{M}_2\,,
\end{eqnarray}
where we have defined the combinations of $\mathfrak{so}(8)$ generators
\begin{eqnarray}
&& \textsf{M}_1 \equiv  \sqrt{3} \textsf{M}_{17} +\textsf{M}_{27} +\textsf{M}_{38} +\textsf{M}_{46} \nonumber \\
&& \textsf{M}_2 \equiv  \sqrt{3} \textsf{M}_{18} -\textsf{M}_{28} +\textsf{M}_{37} +\textsf{M}_{56}  \nonumber \\
&&\textsf{M}_3 \equiv  - 2 \textsf{M}_{23} -\textsf{M}_{45} -\textsf{M}_{78}  \nonumber \\[10pt]
&& \textsf{N}_1 \equiv -\tfrac{1}{\sqrt{3}} \textsf{M}_{15} +\textsf{M}_{25} - \textsf{M}_{34}  + {\tfrac{2}{\sqrt 3}} \textsf{M}_{46} + \textsf{M}_{68} \nonumber \\
&& \textsf{N}_2 \equiv \tfrac{1}{\sqrt{3}} \textsf{M}_{14} +\textsf{M}_{24} + \textsf{M}_{35} + {\tfrac{2}{\sqrt 3}} \textsf{M}_{56} - \textsf{M}_{67}  \nonumber \\
&& \textsf{N}_3 \equiv {\tfrac{2}{\sqrt 3}} \textsf{M}_{16} - {\tfrac{2}{\sqrt 3}} \textsf{M}_{45}  +\textsf{M}_{47} + \textsf{M}_{58}\,.
\end{eqnarray}
These six generators close into the Lie algebra $\mathfrak{so}(4)\approx \mathfrak{su}(2) \oplus \mathfrak{su}(2)$.
This can be seen by taking the further combinations
\begin{eqnarray} \label{Sp1Sp1gen}
 \textsf{J}^{(8)}_x = \tfrac{1}{{4}} ( \textsf{M}_x -\sqrt{3} \textsf{N}_x ) \; , \qquad
 \textsf{L}^{(8)}_x = \tfrac{1}{{4}} ( 3\textsf{M}_x +\sqrt{3} \textsf{N}_x ) \; , \quad x=1,2,3 ,
\end{eqnarray}
and verifying that they satisfy the canonical commutation relations
(\ref{SO4commrel}).
We have thus demonstrated that the spin connection corresponding to the metric (\ref{HMg=0metric}) takes values in $\mathfrak{so}(4)\subset\mathfrak{so}(8)$, which shows that the holonomy of $G_{2(2)}/SO(4)$ is indeed $SO(4) \approx SU(2) \times SU(2)$.
The two $\mathfrak{su}(2)$ components can be seen by writing
\begin{eqnarray} \label{spincon2}
 w =  \omega^x \textsf{J}^{(8)}_x +\Delta^x \textsf{L}^{(8)}_x\,,
\end{eqnarray}
where
we have defined the one-forms $\omega^x$, $\Delta^x$, $x=1,2,3$, as
\begin{equation} \label{Sp1spincon}
 \omega^1 = \sqrt{3} e^5+e^7 \; , \quad  \omega^2 = -\sqrt{3} e^4+e^8 \; , \quad \omega^3 = e^3 -\sqrt{3} e^6 \; , \quad
\end{equation}
and
\begin{equation}
 \Delta^1 = -\tfrac{1}{\sqrt{3}} e^5+e^7 \; , \quad  \Delta^2 = \tfrac{1}{\sqrt{3}} e^4+e^8 \; , \quad \Delta^3 = e^3 + \tfrac{1}{\sqrt{3}} e^6 \; . \quad
\end{equation}
It turns out that the $\omega^x$ are the components of the canonical $Sp(1)$ part of the connection related to
\eqref{quaternion}.
To see this we calculate the curvature of $\omega^x$, defined by
\begin{equation}
-2K^x = d \omega^x +\tfrac{1}{2}\epsilon^{xyz} \omega^y \wedge \omega^z  \, ,
\end{equation}
to find
\begin{eqnarray}  \label{Sp1curv}
K^1 &=& \tfrac12 \left( e^{15} +\sqrt{3} e^{17} -\sqrt{3} e^{25} + e^{27} +\sqrt{3} e^{34}
 + e^{38} - e^{46} -\sqrt{3} e^{68} \right)\,, \nonumber \\
K^2 &=& \tfrac12 \left( -e^{14} +\sqrt{3} e^{18} -\sqrt{3} e^{24} - e^{28} -\sqrt{3} e^{35}
 + e^{37} - e^{56} +\sqrt{3} e^{67} \right)\,, \nonumber \\
K^3 &=& \tfrac12 \left( -2e^{16} - 2 e^{23} + e^{45} -\sqrt{3} e^{47} -\sqrt{3} e^{58} - e^{78} \right)\,,
\end{eqnarray}
where $e^{15} = e^1 \wedge e^5$, etc. Some algebra now
shows that $(J^x)^{\bar u}{}_{\bar v} = \delta^{\bar u \bar w} (K^x)_{\bar w \bar v}$, $x=1,2,3$, is indeed a triplet of complex structures satisfying the quaternion algebra (\ref{quaternion}).
Finally, we note that it can be
checked that the curvature of the $Sp(1) \subset Sp(2)$ connection
$\Delta$ does not lead to a quaternionic structure.

\end{document}